\documentclass[aps,prb,twocolumn,superscriptaddress,showpacs,notitlepage]{revtex4-1}
\pdfoutput=1
\usepackage{color}
\usepackage{amssymb}
\usepackage{amsmath}
\usepackage{appendix}
\usepackage{longtable}
\usepackage{bm}
\usepackage{graphicx}
\usepackage{txfonts}
\usepackage{rotating}

\newcommand{\bea}{\begin{eqnarray}}
\newcommand{\eea}{\end{eqnarray}}
\newcommand{\bei}{\begin{itemize}}
\newcommand{\eei}{\end{itemize}}
\newcommand{\be}{\begin{equation}}
\newcommand{\ee}{\end{equation}}
\newcommand{\bse}{\begin{subequations}}
\newcommand{\ese}{\end{subequations}}
\newcommand{\bfg}{\begin{figure}}
\newcommand{\efg}{\end{figure}}
\newcommand{\eins}{\mbox{$1 \hspace{-1.0mm} {\bf l}$}}

\newcommand{\bese}{\begin{subequations}}
\newcommand{\eese}{\end{subequations}}

\newcommand{\alphatilde}{{\tilde{\alpha}}}

\newcommand{\smequiv}{\! \equiv \!}
\newcommand{\smeq}{\! = \!}

\newcommand{\smap}{\! \approx \!}
\newcommand{\smneq}{\! \neq \!}
\newcommand{\smpl}{\! + \!}

\newcommand{\smmi}{\! - \!}

\newcommand{\ve}{\varepsilon}

\newcommand{\ci}{\mathrm{i}}

\newcommand{\braket}[1]{\left<#1\right>}

\newcommand{\ketLR}[1]{\left|#1\right>}
\newcommand{\ket}[1]{\left|#1\right>}

\newcommand{\braLR}[1]{\left<#1\right|}

\newcommand{\abssqr}[1]{|#1|^{2}}

\makeatletter
\def\Ddots{\mathinner{\mkern1mu\raise\p@
\vbox{\kern7\p@\hbox{.}}\mkern2mu
\raise4\p@\hbox{.}\mkern2mu\raise7\p@\hbox{.}\mkern1mu}}
\makeatother

\begin{document}
\title{Unpaired Floquet Majorana fermions without magnetic fields}
\author{Andres A. Reynoso}
\affiliation{Centre for Engineered Quantum Systems, School of Physics, The University of Sydney, NSW 2006, Australia}
\author{Diego Frustaglia}
\affiliation{Departamento de F\'isica Aplicada II, Universidad de Sevilla, E-41012 Sevilla, Spain}

\begin{abstract}
Quantum wires subject to the combined action of spin-orbit and Zeeman coupling in the presence of \emph{s}-wave pairing potentials (superconducting proximity effect in semiconductors or superfluidity in cold atoms) are one of the most promising systems for the developing of topological phases hosting Majorana fermions.
The breaking of time-reversal symmetry is essential for the appearance of unpaired Majorana fermions. By implementing a
\emph{time-dependent} spin rotation, we show that the standard magnetostatic model maps into a \emph{non-magnetic} one where the breaking of time-reversal symmetry is guaranteed by a periodical change of the spin-orbit coupling axis as a function of time. This suggests the possibility of developing the topological superconducting state of matter driven by external forces in the absence of magnetic fields and magnetic elements. From a practical viewpoint, the scheme avoids the disadvantages of conjugating magnetism and superconductivity, even though the need of a high-frequency driving of spin-orbit coupling may represent a technological challenge. We describe the basic properties of this Floquet system by showing that finite samples host unpaired Majorana fermions at their edges despite the fact that the bulk Floquet quasienergies are gapless and that the Hamiltonian at each instant of time preserves time-reversal symmetry. Remarkably, we identify the mean energy of the Floquet states as a topological indicator. We additionally show that the localized Floquet Majorana fermions are robust under local perturbations. Our results are supported by complementary numerical Floquet simulations.
\end{abstract}

\maketitle

\section{Introduction}

Experimental research in the field of hybrid systems involving spin-orbit coupling (SOC) and superconductivity is currently booming.\cite{LeoAPS2012,*scienceHuntMajoranas2011,*PhysicsAlicea2010,Kouwenhoven2012,RokhinsonMajorana2012,Das2012} It has been predicted that systems based on topological insulators,\cite{FuKane2008,[{The existence of helical edge modes in the topological insulating phase requires the presence of SOC in the bulk, see, for example,~}][{.}]KaneMele2005} Bose-Einstein condensates (BECs) in ultracold atoms,\cite{Chuanwei2008,SatoTakahashi2009prl,Cirac2011prl} or semiconducting quantum wires,\cite{SauDasSarma2010,*Alicea2010prb,*OregRefaelvonOppen2010,PotterLee2010} can realize the topological superconducting phase (TSP). Remarkably, the combination of SOC and proximity to \emph{s}-wave superconductors paves the way for the generation of an effective \emph{p}-wave superconducting pairing. The proposed one-dimensional (1D) samples would host one unpaired\footnote{The term \emph{unpaired} in this paper refers to \emph{single} Majorana states which \emph{do not} appear degenerated--- in energy for static systems or in Floquet quasienergy for time-periodic Hamiltonians--- with other Majorana states localized at the same region.} Majorana fermion at each edge: They effectively behave as systems of 1D spinless paired fermions,\cite{ReadGreen2000} i.e., the Kitaev 1D model.\cite{Kitaev2001} Crucially, in all these physical platforms the access to the TSP is achieved by breaking time-reversal (TR) symmetry with an additional magnetic field, where the corresponding Zeeman energy must overcome a critical value.
Beyond the relevance for fundamental physics, the synthesis of the elusive Majorana fermions (MFs)--- probably realized, already\cite{Kouwenhoven2012,RokhinsonMajorana2012,Das2012,Sasaki2011PRL}--- would unveil a new set of technological possibilities: Since localized MFs are Ising anyons, their topological properties can be profited for applications in quantum information and quantum computation.\cite{Ivanov2001prl,Kitaev2001,Stern2004prb,TQCrmpNayak,AliceaNatPhys2011}

In parallel, topological phases of matter have also been studied in systems out of equilibrium. By applying Floquet theory, it has been shown that time-dependent systems can develop topological phases
that have no analog in static systems.\cite{Kitagawa2010prb}
For example, there have been several studies on graphene subject to electromagnetic radiation in the microwave-THz regime.\cite{Oka2009prb,Rivera2009,FoaCalvoPastawski2011,Zhou2011prb,Busl2012,FertigPRL2011} In the case of circularly polarized radiation, topological insulating features show up giving rise to the existence of gapless edge states. Another interesting proposal is the Floquet topological insulator in semiconducting systems.\cite{FloquetTopoInsNatPhys2011} Moreover, 1D photonic bound states at the interface between two distinct Floquet topological phases have been predicted\cite{Kitagawa2010pra} and detected.\cite{KitagawaWhiteQW2011} Recently, in Ref.~\onlinecite{Cirac2011prl}, localized Floquet Majorana fermions (FMFs) have been predicted in cold-atom quantum wire (with static SOC and magnetic field) due to a time-periodic driving of the chemical potential.\cite{[{More work for this type of FMFs has appeared during the reviewing period of this paper, namely,~}][{~and~}]KunduFMF2013,*LiuBarangerFMF} Similarly, FMF 1D modes were recently predicted to appear at the edges of a cold-atom superfluid 2D system in which the potential of the optical square lattice is periodically modulated.\cite{Liu2012superfluid}

Our starting point here is the 1D model for a quantum wire in which SOC, \emph{s}-wave pairing potential, and Zeeman interaction (all of them static and spatially uniform) coexist.\cite{SauDasSarma2010,*Alicea2010prb,*OregRefaelvonOppen2010}
We gain insight into alternative realizations by applying unitary transformations to such model, where the physical system described by the transformed Hamiltonian can share the topological properties of the original system.\footnote{For
time-independent unitary transformations the topological properties are common to the two linked Hamiltonians. For time-dependent unitary transformations the topological properties are preserved only in very specific cases.}
A successful example of this scheme\cite{Braunecker2010} was reported in Ref.~\onlinecite{MortenKarsten2011}, where a SOC-free topological platform was derived for systems subject to magnetic textures.
We point out that for systems in which magnetic elements are required, it is conceptually useful to be aware of alternatives for replacing their effect.\cite{JakschZoller2003,Mueller2004,Sorensen2005,Lin2009Nature} Inspired by concepts of nuclear magnetic resonance,\cite{Rabi1937} we propose a topological platform for Majorana fermions obtained by mapping the Zeeman term out from the originally static Hamiltonian: A 1D quantum wire subject to a periodically driven SOC, and then the MFs appearing in this context are Floquet Majorana fermions. The proposed system \emph{does not require} any external magnetic field or proximity to magnetic materials avoiding the difficulties of combining superconductivity and magnetism. This profit is counterpointed by the need of a time-periodic modulation of the SOC-axis that should not degrade the pairing potential. However, this non-magnetic platform not only motivates the search and design of physical systems in which the SOC can be changed with time but also is an interesting case study in the active field of topological phases in driven systems.\cite{Kitagawa2011prb,Liu2012superfluid}

By applying the exact mapping between the non-magnetic Floquet system and the magnetic static 1D quantum wire we obtain the effective Hamiltonian for the evolution over one driving period, $T\smeq 2\pi/\Omega$ with $\Omega$ the driving frequency: It is found that above (below) a critical frequency, $\Omega_{\rm c}$, the system is in the Floquet topologically nontrivial  (trivial) superconducting phase. The Floquet quasienergy spectrum for an extended quantum wire consists of a family of excitations forming a Dirac cone that becomes gapless at small wavenumber $k$ only for the critical driving frequency. The system here is gapless in Floquet quasienergies when including solutions for all $k$. However, as pointed out by Kitagawa \emph{et al.} in Ref.~\onlinecite{Kitagawa2010prb}, topological properties can be present under such conditions. We also investigate the mean energies of the Floquet solutions, i.e., the expectation value of the Hamiltonian averaged over the period $T$. At small $k$ we find that the family of solutions in the quasienergy Dirac cone--- closing at a finite quasienergy value--- produces a Dirac cone in mean energy that closes at zero energy for the critical driving frequency.\footnote{Solutions for small $k\neq0$ form a Dirac cone that closes at zero mean energy but the two states at its vertex, $k=0$, have finite mean energies, $\pm\hbar\Omega_{\rm c}/2$.} Remarkably, in this system the topological phase can be distinguished from the trivial phase directly from the mean energy of the Floquet states at small $k$.

We further focus on a finite piece of wire in the Floquet TSP (i.e., $\Omega>\Omega_{\rm c}$), obtaining unpaired FMFs localized at the edges of the sample. This is shown both by using the mapping to the static system and by solving numerically the time-dependent problem. The FMFs appear at a well-defined finite quasienergy, $\pm\hbar\Omega/2$. By working entirely in the time-dependent system we verify the robustness of the FMFs under static local disorder (something expected by virtue of our \emph{exact mapping} to the static system of Ref.~\onlinecite{SauDasSarma2010,*Alicea2010prb,*OregRefaelvonOppen2010}) and investigate different types of interfaces involving Floquet topological systems. Remarkably, the unpaired FMFs appear even though the Hamiltonian preserves TR symmetry at each instant of time. This is possible due to the effective violation of TR symmetry over one driving period. For the sake of completeness, we also explore some examples of SOC drivings preserving TR symmetry finding no FMFs.

The quasienergy at which the FMFs appear agrees with that reported in Ref.~\onlinecite{Cirac2011prl} for one flavor of FMFs. We point out that the mean energies of FMF states \emph{are zero} because their quasienergies are proportional to the driving frequency. Recently, Arimondo \emph{et al.}\cite{Arimondo2012} suggested (with some experimental support) that the statistics of the non-equilibrium populations of the Floquet states could be determined by Bose-Einstein (for bosons) or Fermi-Dirac (for fermions) distributions in the \emph{mean-energy variable}. In that case, the mean energy would play the role of the energy in static systems. Similar conclusions are theoretically drawn in some limits for a Floquet system in contact with a thermal bath.\cite{Breuer2000pre} These results would imply that there might exist some regimes in which the FMFs are the highest mean energy occupied Floquet states because zero energy, in the BdG scheme we are working, lies exactly at the chemical potential of the superconductor: A larger system in thermal equilibrium which is contacted to the Floquet quantum wire.

The paper is organized as follows. In Sec.~\ref{SC:deriving} we outline the derivation of the non-magnetic Floquet platform. In Sec.~\ref{SC:floquet} we explore its properties for both infinite and finite samples, identifying the presence of FMFs. Further discussion and conclusions are presented in Sec.~\ref{SC:conclusions}.

\section{Derivation of two alternative topological platforms}
\label{SC:deriving}

We start by introducing the semiconducting quantum-wire platform of Ref.~\onlinecite{SauDasSarma2010,*Alicea2010prb,*OregRefaelvonOppen2010} in which the interactions are constant as a function of time and position. In the absence of superconductivity, the electrons in the wire follow the Hamiltonian
\bese
\bea
\hat{H}_\mathrm{0,e}= \hat{H}_{\mathrm{kin}} + \hat{H}_{\mathrm{Z}} + \hat{H}_{\mathrm{so}},~~~~~~~~~~~~~~~~ \label{EQ:SOC1}\\
\hat{H}_{\mathrm{kin}}=\frac{p_{x}^{2}}{2 m^*},~~~ \hat{H}_{\mathrm{Z}}=E_{\mathrm{Z}} {\sigma_3},~~
\hat{H}_{\mathrm{so}}=\frac{\alpha}{\hbar} p_x {\sigma_1},~~~~ \label{EQ:SOC}
\eea
\eese
where $m^*$ is the effective electron mass, $E_{\mathrm{Z}}$ is the Zeeman energy, and $\alpha$ is the SOC strength. The $\sigma _{i}$ ($i=1,2,3$) are Pauli matrices operating in spin space, while $E_{\mathrm{Z}}=\frac{1}{2}g \mu_B B_3$ with $B_3$ a magnetic field applied along the direction of $\sigma_3$. We have chosen the most favorable conditions for the development of a TSP, where the SOC axis is orthogonal to the magnetic field. Physically, for a Rashba SOC the $\sigma_1$ axis would be perpendicular to the wire's direction $x$.

Due to a proximity effect induced by a nearby \emph{s}-wave bulk superconductor there is a nonzero electron-hole pairing characterized by the energy gap $\Delta_0=|\Delta|$. Electrostatic gates control the chemical potential $\mu$. The system is  described by the Bogoliubov-deGennes (BdG) equation\cite{deGennesBook}
\bese
\begin{eqnarray}
\label{EQ:H0}
\mathcal{H}_0=\int \bar{\Psi}^\dagger(x) \hat{H}_0 \bar{\Psi}(x) dx ,~~  \bar{\Psi}^\dagger=\left(\hat{\Psi}^\dagger_\uparrow,\hat{\Psi}^\dagger_\downarrow,\hat{\Psi}_\downarrow,-\hat{\Psi}_\uparrow \right)\,, ~~~~~~~\label{EQ:bdgeq}\\
\hat{H}_0 \ketLR{\phi(t)}=\ci \hbar \frac{d}{d t}  \ket{\phi(t)}\,,~~~~~~~~~~~~~~~~~~~~~~
\\
\hat{H}_0=\left(\begin{array}{cc} \hat{H}_{0,\mathrm{e}}-\mu & \Delta\\ \Delta^*  &  \mu - \hat{H}_{0,\mathrm{h}} \end{array} \right),~~~~~~~~~~~~~~~~~~~~~
\label{EQ:HBdG}
\end{eqnarray}
\label{EQ:HBdGfull}
\eese
where $\hat{\Psi}_{\uparrow,(\downarrow)}(x)$ is the annihilation operator
for electrons with up (down) spin at position $x$ and time $t$. In Eq.~\eqref{EQ:HBdG}, the quantities $\mu$, $\Delta$ and $\Delta^*$ are to be interpreted as multiplied by the identity $\eins_{2\times2}$. The Hamiltonian for the holes, $\mu-\hat{H}_{0,\mathrm{h}}$, is obtained by time reversing $\hat{H}_{0,\mathrm{e}}$,
\be
\hat{H}_{0,\mathrm{h}}= \mathcal{T}^{-1} {\hat{H}}_{0,\mathrm{e}} \mathcal{T}= \sigma_2 {\hat{H}}_{0,\mathrm{e}}^* \sigma_2.
\label{EQ:TRS}
\ee
where $\mathcal{T}\smeq -\ci\sigma_2 K$ is the TR operator with $K$ the complex conjugation. The $\hat{H}_{\mathrm{kin}}$ and $\hat{H}_{\mathrm{so}}$ are terms preserving TR symmetry in $\hat{H}_{0,\mathrm{e}}$, while the Zeeman contribution, $\hat{H}_{\mathrm{Z}}$, reverses its sign under TR.

In Eq.\eqref{EQ:HBdG}, for simplifying treatment on next sections we have written the BdG-Schroedinger equation operating on a ket, $\ket{\phi(t)}$. Indeed, including the case of a BdG time-dependent Hamiltonian, $\hat{H}'(t)$,  the most natural representation of those states is a Nambu spinor, ${\it\bar{\Phi}}(x,t)$ which is written in the basis of $\bar{\Psi}(x,t)$ and evolves satisfying the BdG-Schroedinger equation $\left(\hat{H}'(t) \smmi \ci \hbar \frac{\partial}{\partial t}\right)   {\it\bar{\Phi}}(x,t)\smeq 0$. The Nambu field operator allows one to write the fermionic annihilation and creation operator associated with any solution ${\it\bar{\Phi}}(x,t)$, this will be useful below; In particular the annihilation operator written in the Schroedinger and Heisenberg pictures are
\bese
\bea
\hat{\Phi}_S&=& \int{\it\bar{\Phi}}^\dagger(x,0) \cdot\bar{\Psi}(x)  dx =\hat{\Phi}_H(0), \label{EQ:annihilS}\\
\hat{\Phi}_H(t)&=& \hat{\mathcal{U}'}^\dagger(t,0) \hat{\Phi}_S  \hat{\mathcal{U}'}(t,0) ~,
\label{EQ:annihilH}
\eea
\label{EQ:annihil}
\eese
where $\hat{\mathcal{U}'}(t,0)$ is the evolution operator from time $0$ to $t$.

The Hamiltonian $\hat{H}_0$ of Eq.~(\ref{EQ:HBdGfull}) has been studied extensively. It is known that the system is in the TSP for Zeeman energies above a critical value,
\be
|E_{\mathrm{Z}}|> E_\mathrm{Z}^{\rm c}\equiv\sqrt{\Delta_0^2 +\mu^2}.
\label{EQ:condition}
\ee
This can be seen by solving $\hat{H}_{0}  {\it\bar{\Phi}}=E {\it\bar{\Phi}}$ for an infinite wire and calculating the $Z_2$ topological invariant that distinguishes both gapped phases.\cite{Kitaev2001} A topological phase transition involves a closing and reopening of the band gap as a function some parameter.\cite{topoinsulators2011rmp} Here, the gap closes when $|E_{\mathrm{Z}}|\smeq E^{\rm c}_\mathrm{Z}$. We get insight into the relevant states around the critical condition by working at Zeeman energies, $E_{\mathrm{Z}}=E_\mathrm{Z}^{\rm c}+\Delta e_{\mathrm{Z}}$.
We focus on the solutions for small linear momentum $p_x\smeq\hbar k$ (the solutions for $|k|\smap k_\mathrm{F}$ are gapped due to the \emph{s}-wave superconducting pairing). Close to $E\smeq 0$, the solutions organize in two branches which, at the leading order in the wavenumber $k$, read
\be
E^0_\pm(k) = \pm \sqrt{\left(\alpha k \sin{\phi_0} \right)^2+\Delta e_{\mathrm{Z}}^2},
\ee
with $\phi_0 \equiv \arctan \frac{\Delta_0}{-\mu}$. These solutions are gapless only at the critical Zeeman field, forming a Dirac cone
\be
E^0_\pm(k) = \pm \alpha k \sin \phi_0 +O\left(k^2\right) = \pm\alpha k  \frac{\Delta_0}{\sqrt{\Delta_0^2+\mu^2}}+O\left(k^2\right).
\ee
At $k\smneq0$, these branches correspond to states that are the symmetric and antisymmetric combination of states unperturbed by SOC, with energies $\left.E^0_{1,2}(k)\right|_{\alpha\smeq 0}\smeq\pm \Delta e_{\mathrm{Z}}\smpl O\left(k^2\right)$. The unperturbed states are ${\it\bar{\Phi}}^0_1\smeq\left(\cos{\frac{\phi_0}{2}},0,\sin{\frac{\phi_0}{2}},0\right)^\dagger \frac{{\rm e}^{\ci k x}}{\sqrt{L}}$ and ${\it\bar{\Phi}}^0_2\smeq\left(0,-\sin{\frac{\phi_0}{2}},0,\cos{\frac{\phi_0}{2}}\right)^\dagger \frac{{\rm e}^{\ci k x}}{\sqrt{L}}$, where the Nambu-space spin axis is taken along $\sigma_3$.

Importantly, for $\Delta e_{\mathrm{Z}}>0$ ($\Delta e_{\mathrm{Z}}<0$) the solution at $k\smeq 0$ with $E^0>0$ is ${\it\bar{\Phi}}^0_1$ (${\it\bar{\Phi}}^0_2$).
An exchange of these eigenstates (i.e., moving from one gapped phase to the other) can only be realized after a band inversion: Right at the critical condition the gap closes and both solutions become degenerated. It is then impossible to go from one situation to the other by a smooth perturbation of the Hamiltonian without closing and reopening the gap. This indicates that the two gapped phases are topologically distinct.\cite{topoinsulators2011rmp}
The topologically trivial phase is defined by $E_\mathrm{Z}<\sqrt{\Delta_0^2 +\mu^2}$ since this situation includes the vacuum phase characterized by a large chemical potential $\mu<0$.
Therefore, gapless states only appear at the interfaces of topological regions (where $E_\mathrm{Z}>\sqrt{\Delta_0^2 +\mu^2}~$) with non-topological ones (as, e.g., the vacuum).

This class of midgap states is well-known to appear in Dirac-like equations as a consequence of a sign change in the mass term.\cite{JackiwRebbi1976} For the BdG equation, due to particle-hole symmetry, for each eigenstate with $+E_a$ created by the operator $\hat{\Phi}_a^\dagger$ (taken either in the Heisenberg or the Schroedinger pictures, see Eq.\eqref{EQ:annihil}) there is another one at $-E_a$ which is created by $\hat{\Phi}_a$.\cite{FlensbergMajoranaReview2012} Therefore, if a physical system has an eigenstate satisfying the reality condition $\hat{\Phi}_a\smeq \hat{\Phi}_a^\dagger$ (i.e., being its own antiparticle), it must appear at $E_a\smeq 0$. Furthermore, the zero energy state is protected from local perturbations by an energy gap\cite{SauDasSarma2010,*Alicea2010prb,*OregRefaelvonOppen2010} when it is found unpaired: Localized in space and free of any overlapping with other Majorana solutions. Notice that as real fermions have spin nonzero, effective spinless fermions can only arise after breaking TR symmetry.\footnote{Such breaking can be simple (e.g., induced by an external magnetic field) or may spontaneously develop due to interactions.} Otherwise, the presence of a nontrivial TR symmetry (one which is not represented by the identity operator when applied to bound states) would lead to two-fold degeneracy of bound states.\cite{FlensbergMajoranaReview2012} For the Majorana fermions discussed here the system would host more than one bound Majorana fermion at each edge of the sample. Here, for quantum wires subject to static SOC and Zeeman fields in the TSP, the required TR-symmetry breaking condition is obviously satisfied.

\subsection{SOC-free magnetostatic platform}

Consider the joint kinetic and SOC terms in Eqs.~(\ref{EQ:SOC1}) and (\ref{EQ:SOC}), $\hat{H}_{\mathrm{kin}}+\hat{H}_{\mathrm{so}}\smeq\frac{1}{2m^*}(p_x +\hbar k_\mathrm{so}\sigma_1)^2 - E_\mathrm{so}$, with $k_\mathrm{so} \smequiv \frac{m^*\alpha}{\hbar^2}$ and $E_\mathrm{so}\smequiv \frac{\hbar^2  k_\mathrm{so}^2}{2 m^*}\smeq \frac{\alpha^2 m^*}{2 \hbar^2}$. A spin-dependent shift in momentum,\cite{AleinerFalko2001} $\hat{U}_\mathrm{1}\smeq\exp\left(-\ci  k_\mathrm{so} x \sigma_1\right)$, cancels the SOC term. Since $\hat{U}_\mathrm{1}$ rotates the spin by an angle $2 k_\mathrm{so} x$ around the $\sigma_1$-axis, the Zeeman term--- constant in the original Hamiltonian--- changes its axis in the (2,3) plane as a function of position. The transformed Hamiltonian, $\hat{H}_\mathrm{1,e} \smeq \hat{U}_\mathrm{1}^\dagger \hat{H}_{0,\mathrm{e}} \hat{U}_\mathrm{1}$, becomes,\cite{Braunecker2010}
\be
\label{Ha}
\hat{H}_\mathrm{1,e} =\frac{p_x^2}{2m^*}  - E_\mathrm{so} + E_{\mathrm{Z}}\left[ {\cos \left(2k_\mathrm{so} x\right) \sigma_3 + \sin \left(2 k_\mathrm{so} x\right) \sigma_2}\right]\,.
\ee
Recently, Kjaergaard \emph{et al.}\cite{MortenKarsten2011} demonstrated that a system of electrons following Hamiltonian $\hat{H}_\mathrm{1,e}$ in the presence of a \emph{s}-wave superconducting pairing potential--- see Eq.\eqref{EQ:HBdGfull}--- can develop a TSP. This SOC-less platform could be realized for instance by an engineered array of nearby micromagnets.

A direct comparison with $\hat{H}_\mathrm{0,e}$ indicates that the amplitude of the rotating magnetic field must be sufficiently \emph{strong} to fulfill the condition of Eq.\eqref{EQ:condition} (the chemical potential in the equation must be replaced by $\mu-E_\mathrm{so}$ because the energy shift $E_\mathrm{so}$ is absent in this zero-SOC  platform). The results of Ref.~\onlinecite{MortenKarsten2011} support the approach adopted here for exploring new physical platforms for TSPs by unitarily removing a particular interacting term from an  known topological Hamiltonian. Remarkably, this approach allows the identification of topological platforms even in situations where the mapping is not exact. This was shown numerically in Ref. \onlinecite{MortenKarsten2011} for non-sinusoidal magnetic textures in the absence of SOC.

It is interesting to discuss Eq.~(\ref{Ha}) from the point of view of the electron spin dynamics ($\Delta_0\smeq0$).
The relevant parameters are: The characteristic length $L=2\pi k_\mathrm{so}^{-1}$ over which the magnetic texture suffers a significant change, the magnetic field strength $E_Z$, and the spin carrier Fermi velocity $v_{\rm F}$.
For a given $L$, the effective SOC becomes significant for relatively weak textures (moderate $E_Z$). This corresponds to the regime of  \emph{non-adiabatic} spin transport, where the spin eigenstates of Eq.~(\ref{Ha}) are prevented from being fully aligned with the local magnetic field. (For a given strength $E_Z$, instead, the same regime can be achieved for a relatively small $L$). Otherwise, no spin mixing survives from the magnetic texture: In the TSP this would minimize the effective gap protecting the MFs. More accurately, the \emph{adiabatic} regime is defined in the limit $\omega_{\rm s} \gg  2\pi/t_{\rm c}$, where $\omega_{\rm s}=E_{\rm Z}/\hbar$ is the Larmor frequency of spin precession and $t_{\rm c}=L/v_{\rm F}$ is the time
it takes the spin carriers to cover the length $L$ (see Refs.~\onlinecite{FrustagliaRichter2001}~and~\onlinecite{PoppFrustaglia2003} for a detailed discussion). Non-adiabatic dynamics--- i.e., a significant gap protecting the MFs--- requires $\omega_{\rm s} \sim  2\pi/t_{\rm c}$.

\subsection{Non-magnetic Floquet platform}
\label{SC:Urt}

Here we get rid of the Zeeman term of Eq.~(\ref{EQ:SOC1}) by applying a global time-dependent rotation to the spins along the magnetic-field axis with an appropriate frequency. For simplicity, we first consider the electron block of the Hamiltonian (not the BdG equation) where the mentioned transformation reduces to $\hat{U}_\mathrm{r}(t)\smequiv\exp\left(-\ci \sigma_3 t E_\mathrm{Z}/\hbar \right)$. The time evolution of the rotated states
\be
\ketLR{\phi^{\mathrm{rot}}(t)}\smequiv \hat{U}^\dagger_\mathrm{r}(t)\ketLR{\phi(t)\vphantom{\phi^{\mathrm{rot}}}}\,,
\ee
is given by the equation
$\ci \hbar \frac{d}{dt}  \ketLR{\phi^{\mathrm{rot}}(t)} \smeq \hat{H}_\mathrm{e}(t) \ketLR{\phi^{\mathrm{rot}}(t)}$, with
\bea
\hat{H}_\mathrm{e}(t)&\equiv& \hat{U}^\dagger_\mathrm{r}(t) \hat{H}_\mathrm{0,e} \hat{U}_\mathrm{r}(t)- \ci \hbar \hat{U}^\dagger_\mathrm{r}(t) \frac{d \hat{U}_\mathrm{r}(t)}{dt} \nonumber \\&=& \hat{U}^\dagger_\mathrm{r}(t) \hat{H}_\mathrm{0,e} \hat{U}_\mathrm{r}(t)-E_\mathrm{Z}\sigma_3 \nonumber \\
&=& \frac{p_x^2}{2m^*}  + \frac{\alpha}{\hbar} p_x \left[ {\cos \left(2\frac{E_\mathrm{Z}}{\hbar} t\right) \sigma_1 -\sin \left(2\frac{E_\mathrm{Z}}{\hbar}t\right) \sigma_2}\right],~~~~
\label{EQ:tx}
\eea
where the equation of motion for the original states is $\ci\hbar \frac{d}{dt}  \ketLR{\phi(t)} \smeq \hat{H}_\mathrm{0,e}(t) \ketLR{\phi(t)}$. Therefore, if a physical system follows the time-dependent $\hat{H}_\mathrm{e}(t)$ \emph{in the Schroedinger picture} it can be mapped into the static
$\hat{H}_\mathrm{0,e}$ by the unitary transformation $\hat{U}_\mathrm{r}(t)$.

Assuming that the system of electrons described by $\hat{H}_\mathrm{0,e}(t)$ is in proximity with and \emph{s}-wave superconductor one needs the instantaneous time-reversal operator of the electrons, $\hat{H}_{\mathrm{h}}(t)\smeq \mathcal{T}^{-1} {\hat{H}}_{\mathrm{e}} (t)\mathcal{T}$. The latter enters as the block $\mu\smmi\hat{H}_{\mathrm{h}}(t)$ in a time-dependent version of the BdG equation presented in Eq.\eqref{EQ:HBdGfull}. In this case it can be shown that $\hat{H}_\mathrm{h}(t)\smeq \hat{U}^\dagger_\mathrm{r}(t) \hat{H}_\mathrm{0,h} \hat{U}_\mathrm{r}(t)\smpl E_\mathrm{Z}\sigma_3$: The magnetic field of the hole sector in the static model appears canceled in the transformed hole system. Importantly, the superconducting pairing terms of the BdG Hamiltonian transform trivially under the full Nambu space transformation (see the definition of $\hat{U}_\mathrm{R}(t)$ in Eq.\eqref{EQ:UR} below).

It is clear that the excitations for the systems described by either $\hat{H}_\mathrm{e}(t)$ or $\hat{H}_\mathrm{0,e}$ (both assumed to be in the Schroedinger picture) are not identical. Even in absence of spin-orbit coupling, where the $\hat{H}_\mathrm{e}(t)$ is \emph{time independent}, both Hamiltonians have different energy spectra. However, assuming electron-hole \emph{s}-wave pairing, it can be shown that both systems share the topologically trivial phase. Similarly, as we show, for the non-magnetic system with a periodically rotating spin-orbit coupling axis our mapping allows to predict the existence of a Floquet superconducting topological phase.

Here, the reason why topological properties are also expected in the time-dependent system lies, actually, in the simplicity of the linking transformation. In the general case, on the other hand, the solutions of \emph{any} BdG Hamiltonian $\hat{H}_{\Theta}(t)$--- including Hamiltonians without topological properties--- can be unitarily mapped into the solutions of $\hat{H}_{0}$ given in Eq.\eqref{EQ:HBdGfull} (the one based on $\hat{H}_{0,\mathrm{e}}$). This is possible provided the unitary transformation is sufficiently complicated to the point of introducing the requirements of the topological phase. More explicitly, starting from the solutions to the arbitrary Hamiltonian, $(\hat{H}_{\Theta}(t)-\ci\hbar\frac{d}{dt}) \ketLR{\phi^\Theta(t)}=0$, the topological Hamiltonian $\hat{H}_{0}$ governs the time dependence of the transformed states if and only if $\ketLR{\phi^{\vphantom{\Theta}\mathrm{trans}}(t)}\smeq \hat{\mathcal{U}}_\mathrm{top}(t,0)\hat{\mathcal{U}}^\dagger_{\Theta}(t,0)\ketLR{\phi^\Theta(t)}$, where $\hat{\mathcal{U}}_\Theta(t,0)$ and $\hat{\mathcal{U}}_\mathrm{top}(t,0)$ are the time-evolution operators from 0 to $t$ associated with $\hat{H}_\Theta$ and $\hat{H}_{0}$, respectively. In our case, $\hat{H}_\Theta\smeq \hat{H}(t)\smeq\hat{U}^\dagger_\mathrm{R}(t)\hat{H}_{0}\hat{U}_\mathrm{R}(t)\smmi\ci \hbar \hat{U}^\dagger_\mathrm{R}(t) \frac{d \hat{U}_\mathrm{R}(t)}{dt}$ [which is the BdG Hamiltonian based on $\hat{H}_\mathrm{e}(t)$ in presence of a pairing potential, see Eq.\eqref{EQ:HBdGfulltd}], where
\bea
\hat{U}_\mathrm{R}(t) &\equiv& \hat{\mathcal{U}}_\mathrm{top}(t,0)\hat{\mathcal{U}}^\dagger_\Theta(t,0)=\left(\begin{array}{cc}
\hat{{U}}_\mathrm{r}(t)&0\\
0&\hat{{U}}_\mathrm{r}(t)
\end{array}\right)\nonumber\\&=&\left(\begin{array}{cccc}
{\mathrm{e}}^{-\frac{\ci}{\hbar}E_\mathrm{Z} t }&0&0&0\\
0&{\mathrm{e}}^{\frac{\ci}{\hbar}E_\mathrm{Z} t }&0&0\\
0&0&{\mathrm{e}}^{-\frac{\ci}{\hbar}E_\mathrm{Z} t }&0\\
0&0&0&{\mathrm{e}}^{\frac{\ci}{\hbar}E_\mathrm{Z} t }
\end{array}\right),
\label{EQ:UR}
\eea
is a rotating-frame spin transformation that does not \emph{introduce} topological superconducting features. Therefore, topological properties in either system, if any, must be linked. In the following section, after introducing the Floquet formalism, we show that
the solutions of the non-magnetic $\mathrm{BdG}$ Hamiltonian $\hat{H}(t)$ present topological features related to those of $\hat{H}_{0}$.
This is because $\hat{U}_\mathrm{R}(t)$ establishes a very simple mapping between the energy spectrum of the static $\hat{H}_{0}$ and the Floquet quasienergies of $\hat{H}(t)$. Furthermore, we see that each eigenstate of $\hat{H}_{0}$ is associated with one Floquet quasi-energy state sharing its properties.

\section{Properties of non-magnetic periodically driven system}
\label{SC:floquet}

We consider a system of electrons described by the time-dependent Hamiltonian of Eq.~\eqref{EQ:tx} in the Schroedinger picture. The rotating SOC coexists with a constant \emph{s}-wave pairing potential, where the
corresponding BdG equation reads
\bese
\begin{eqnarray}
\label{EQ:H0td}
\mathcal{H}(t)&=&\int \bar{\Psi}^\dagger(x) \hat{H}(t) \bar{\Psi}(x) dx ,\\
\hat{H}(t)&=&\left(\begin{array}{cc} \hat{H}_{\mathrm{e}}(t)-\mu & \Delta\\ \Delta^*  &  \mu - \hat{H}_{\mathrm{h}}(t) \end{array} \right).~~~~~~~~~~~~~~~~~~~~~
\label{EQ:HBdGtd}
\end{eqnarray}
\label{EQ:HBdGfulltd}
\eese
Without loss of generality, we choose a superconducting phase factor equal to zero such that $\Delta=\Delta_0$. We introduce a driving frequency $\Omega$ for the time-dependent component of the SOC axis, $\mathbf{\Lambda}(t)$, including also an additional static contribution, $\mathbf{\Lambda}^{(0)}$. Using these definitions, we write the electron and hole Hamiltonian as
\be
\hat{H}_\mathrm{e}(t)=\hat{H}_\mathrm{h}(t)= \frac{p_x^2}{2m^*}  + \frac{p_x}{\hbar} \left( \mathbf{\Lambda}(t)+ \mathbf{\Lambda}^{(0)} \right) \cdot \boldsymbol{\sigma},
\label{EQ:HeHhTime}
\ee
where $\boldsymbol{\sigma}\smeq(\sigma_1,\sigma_2,\sigma_3)$. Equation \eqref{EQ:HeHhTime} follows from Eq.\eqref{EQ:TRS} after noticing that, for any given time $t'$, $\hat{H}_\mathrm{e}(t')$ is a time-reversal symmetric operator. For the ideal rotating (IR) case--- which is the best candidate for hosting Floquet Majorana fermions--- the SOC vectors are,
\bese
\bea
\mathbf{\Lambda}(t)&=&\mathbf{\Lambda}_\mathrm{IR}(t) \equiv \left(\alphatilde\cos \left(\Omega t\right),-\alphatilde\sin \left(\Omega t\right),0\right),\\
\mathbf{\Lambda}^{(0)}&=&\mathbf{0} \equiv (0,0,0).
\eea
\label{EQ:ideal}
\eese
Notwithstanding, we also discuss some scenarios away from this ideal situation.

\subsubsection*{Periodically driven Schroedinger equation and Floquet systems}

As discussed in Sec.\ref{SC:Urt}, 1D systems subject to a rotating SOC are linked to the magnetostatic, topological quantum wire. Time-dependent periodic systems are better described in the context of the Floquet theory, which we summarize in the following (further details are presented in the Appendix \ref{AP:A}). The starting point is the need to solve the Schroedinger equation,
\be
\left(\hat{H}(t)-\ci\hbar \frac{d}{dt}\right)\ketLR{\phi(t)}=0,
\label{EQ:eqSch}
\ee
for a periodically driven Hamiltonian, $\hat{H}(t)\smeq\hat{H}(t\smpl T)$. Energy is not conserved. However, by virtue of the Floquet theorem the solutions can be written as
\be
\ketLR{\phi^{\vphantom{T}}_{a}(t)} = {\rm e}^{-\frac{\ci}{\hbar} \varepsilon_{a} t}\ketLR{\phi^T_{a}(t)},
\label{EQ:physical}
\ee
where $T\smeq \frac{2\pi}{\Omega}$ is the driving period, $\ketLR{\phi^T_{a}(t)}$ is a periodic state, and the subindex $a$ encodes the quantum numbers of the different solutions. We define the Floquet operator $H_{\rm F}\smequiv \left(\hat{H}(t)-\ci\hbar \frac{d}{dt}\right)$. As $\ketLR{\phi_{a}(t)}$ is a solution of the Schroedinger equation,  it holds $H_F\ketLR{\phi_{a}(t)}=0$. One then finds
\be
\left(\hat{H}(t)-\ci\hbar \frac{d}{dt}\right) \ketLR{\phi^T_{a}(t)}=\varepsilon_{a}\ketLR{\phi^T_{a}(t)}.
\label{EQ:floquetQES}
\ee
The quasienergies, $\varepsilon_a$, are the eigenvalues of the Floquet operator $H_{\rm F}$. The corresponding eigenvectors are the Floquet quasienergy states (QESs), $\ketLR{\phi^T_{a}(t)}$. Besides, the evolution operator associated with $\hat{H}(t)$ has the form $\hat{\mathcal{U}}\left(t,t_0\right)\smeq {\bm{T}}_t\exp{\left(-\frac{\ci}{\hbar} \! \int_{t_0}^t  \hat{H}(t') dt' \right)}$, where ${\bm{T}}_t$ stands for time ordering. From Eq.\eqref{EQ:floquetQES} it then follows
\be
\hat{\mathcal{U}}\left(t_0+T,t_0\right)  \ketLR{\phi^T_{a}(t_0)}=  {\rm e}^{-\frac{\ci}{\hbar}\ve_a T} \ketLR{\phi^T_{a}(t_0)}.
\label{EQ:evolOp}
\ee
Hence, the quasienergies can also be extracted from the phase factors ${\rm e}^{-\frac{\ci}{\hbar}\ve_a T}$ (the eigenvalues of the evolution operator over one driving period).

\subsubsection*{Mean energy of Floquet quasienergy states}

The mean energy of a Floquet QES is defined as the expectation value of the Hamiltonian averaged over one driving period:
\bea
\overline{E}_a&=&\frac{1}{T}\int_0^T dt\braket{\phi^T_{a}(t)|\hat{H}(t)|\phi^T_{a}(t)}\nonumber \\
&=&\frac{1}{T}\int_0^T dt\braket{\phi^T_{a}(t)\left|\hat{H}(t)-\ci \hbar \frac{d}{dt}+\ci \hbar \frac{d}{dt}\right|\phi^T_{a}(t)}\nonumber
\\&=&\varepsilon_a -  \Omega \frac{\partial \varepsilon_a}{\partial\Omega},
\label{EQ:meanEnergy}
\eea
where we use an extension of the Hellmann-Feynman theorem (for the Floquet operator $H_{\rm F}$).\cite{Fainshtein1978} Notice that the set of quasienergies $\ve_a$ for a given frequency $\Omega$--- as usually presented in most numerical treatments of Floquet systems--- does not provide information on the associated $\overline{E}_a$. Obtaining the mean energies requires either the derivative $\partial\varepsilon_{a}/\partial\Omega$ or the periodic eigenstate $\ket{\phi^T_{a}(t)}$ in order to evaluate Eq.~(\ref{EQ:meanEnergy}).

Starting from a Floquet state $\ket{\phi^T_{a}(t)}$ with quasienergy $\ve_a$, we find that the mean energy is identical for all associated shifted states of Eq.~(\ref{EQ:shifted}) with quasienergy $\varepsilon_{a} + n \hbar \Omega$, since
\be
\overline{E}^{n\textrm{-shift}}_a=(\varepsilon_{a} + n \hbar \Omega)-\Omega\frac{\partial}{\partial \Omega}\left(\varepsilon_{a} + n \hbar \Omega\right) =\overline{E}_a.
\ee
In summary, the mean energy is a useful quantity for classifying the Floquet QESs and the physical state $\ket{\phi_{a}(t)}$ of Eq.~(\ref{EQ:physical}). Furthermore, as mentioned above, this quantity may be relevant for determining the occupancies of Floquet states.\cite{Arimondo2012,Breuer2000pre} An important question addressed in the following is how the mean energies of Floquet states behave at both sides of the topological transition.

\subsection{Bulk properties}
\label{SC:infinite}

We now investigate the Floquet solutions of the time-dependent system in its infinitely long version. For arbitrary periodic driving, one must proceed numerically because complexity impedes analytical treatments. For the ideal rotating case our numerical implementation---see Appendix \ref{AP:A}---is in full agreement with the analytical results that, as we show in detail below, follow from the mapping to the static topological problem discussed in Sec.\ref{SC:Urt}.

The linear momentum, $p_x\smeq \hbar k$, is a good quantum number as the system is translationally invariant. For each wavenumber $k$ the Hamiltonian $\hat{H}_k(t)$ is a $4\times4$ matrix operating in Nambu space. The trial state $\ketLR{\phi_{k}(t)} \equiv \ketLR{k} \ketLR{\varphi_{k}(t)}$ in the Schroedinger equation \eqref{EQ:eqSch} leads to a differential equation for the Nambu ket, $\ketLR{\varphi_{k}(t)}$:
\be
\left(\hat{H}_k(t)-\ci\hbar \frac{d}{dt}\right) \ketLR{\varphi_{k}(t)}=H_{\rm F}^k~\ketLR{\varphi_{k}(t)} =0.
\label{EQ:floquetQESk}
\ee
For each value of $k$ there exist four solutions. We label them as $\ketLR{\varphi_{k,a}(t)}$, with $a=1,\dots,4$. The knowledge of $\ketLR{\varphi_{k,a}(t)}$ allows us to write the solutions as in Eq.\eqref{EQ:physical} by virtue of the Floquet theorem for the periodically driven $\hat{H}_k(t)$.

\begin{figure}[!t]\begin{center}
\includegraphics[width=0.48\textwidth]{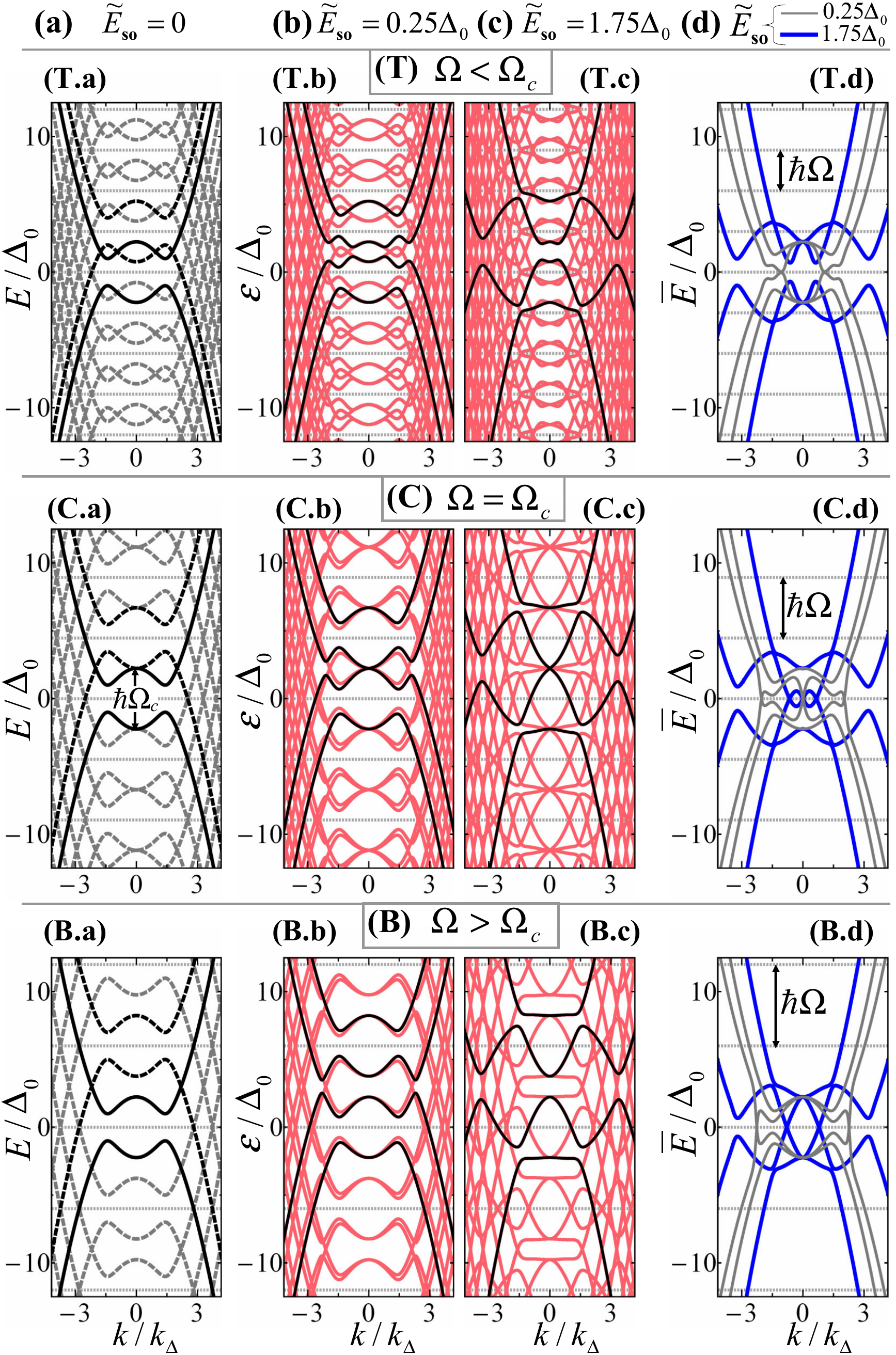}
\vspace{-0.6cm}
\caption{\label{FG:fig1} Ideal rotating (IR) case of Eq.\eqref{EQ:ideal}: Floquet quasienergies and mean energies.
Reference energy and momentum are $\Delta_0$ and $k_\Delta\smeq \sqrt{2 m^* \Delta_0} /\hbar$, respectively.  We set $\mu\smeq 2\Delta_0$ and thus $\hbar\Omega_{\rm c}=2\sqrt{5}\Delta_0$. Top (T), center (C) and bottom (B) panels show results for $\Omega=0.9\Omega_{\rm c}$, $\Omega=\Omega_{\rm c}$ and $\Omega=1.1\Omega_{\rm c}$, respectively. Dotted horizontal lines at $n\hbar\Omega$ are included in all panels. Panels (a) show the spin-degenerated dispersion relations $E_{\pm,k\sigma}$ of Eq.~(\ref{EQ:NakedEnergies}) in the absence of the IR SOC-field (solid lines). Dashed lines correspond to $E_{\pm,k\sigma}\smpl m \hbar\Omega$ branches which are important for the mixing due to the periodic driving, see Eq.\eqref{EQ:floqMatrix}. Panels (b) and (c) present Floquet quasienergies $\varepsilon$  for finite IR SOC-field of intensity $\tilde{E}_\mathrm{so} \equiv \alphatilde^2 m^*/(2 \hbar^2) = 0.25\Delta_0$ and  $1.75\Delta_0$, respectively. Numerical results show an excellent agreement with analytical calculations obtained by mapping the time-dependent problem to the static system having a constant SOC and a finite Zeeman field: The solid black lines depict the four bands of the corresponding static system (shifted in energy by $+\hbar\Omega/2$, see text). Panels (d) show the mean energies $\overline{E}$ of the Floquet states for the simulated IR SOC-field amplitudes. In (C.b) and (C.c), two branches of Floquet solutions close in a Dirac-like quasienergy cone for small $k$. The reopening of the Dirac cone for $\Omega>\Omega_{\rm c}$, (B.b) and (B.c), indicates a topological phase distinct from that for  $\Omega<\Omega_{\rm c}$. The mean energy shows a distinct pattern on each phase (see text and Fig.\ref{FG:fig2}).}
\end{center}
\vspace{-0.6cm}
\end{figure}

\subsubsection{The IR driving: Floquet bands and Majorana states.}

Figure \ref{FG:fig1} shows numerical results for the time-dependent SOC of Eq.\eqref{EQ:ideal}, where the SOC axis rotates harmonically around the $\sigma_3$ axis. As a starting point, in Figs. \ref{FG:fig1}(T.a), \ref{FG:fig1}(C.a) and \ref{FG:fig1}(B.a), we show the dispersion in the absence of driving, corresponding to the eigenenergies $E(k)$ of the static part of the Hamiltonian, $\hat{H}^{(0)}$--- see Eq.~(\ref{EQ:HtFourier}). The only interacting term in $\hat{H}^{(0)}$ is the pairing potential, electron-hole states are spin-degenerated with eigenenergies
\be
E_{\pm,k\sigma}=\pm \sqrt{\Delta_0^2+ \left(\hbar^2k^2/(2m^*)-\mu\right)^2}.
\label{EQ:NakedEnergies}
\ee
These appear as solid lines. Additionally, the same dispersions shifted by an integer multiple of $\hbar\Omega$
are displayed as dashed lines. Those bands are relevant for the time-dependent system since, as shown in Eq.\eqref{EQ:floqMatrix}, the operators $\hat{H}^{(m)}$ of the Fourier decomposition of the full Hamiltonian (see Eq.~(\ref{EQ:HtFourier}) with $m\smneq 0$) mix solutions of $\hat{H}^{(0)}$ with energies differing in $m\hbar\Omega$.

The Fourier decomposition of the interacting term $p_x  \mathbf{\Lambda}_\mathrm{IR}(t) \cdot \boldsymbol{\sigma}/\hbar$ appearing in both $\hat{H}_\mathrm{e}(t)$ and $\hat{H}_\mathrm{h}(t)$, with $\mathbf{\Lambda}_\mathrm{IR}(t)$ given in Eq.\eqref{EQ:ideal}, determines that the only off-diagonal terms contributing to Eq.~(\ref{EQ:floqMatrix}) are
\be
\hat{H}^{(\mp 1)}_\mathrm{e}=\hat{H}^{(\mp 1)}_\mathrm{h} = \frac{\alphatilde}{\hbar} p_x \sigma_\pm,
\label{EQ:HFourier1}
\ee
where $\sigma_\pm=\sigma_1\pm\ci\sigma_2$ are the $\sigma_3$-spin raising and lowering operators. This means that the driving introduces mixing between $\hat{H}^{(0)}\smpl m\hbar\Omega$ blocks with $(m\smmi m')\smeq 1$. For each $k$ one needs to solve a time-independent Floquet equation as Eq.\eqref{EQ:floqMatrix}, where each block is a $4\times 4$ matrix. The amplitude of the IR mixing terms is $\pm\alphatilde k$. This means that at $k\smeq 0$ the system is static and the Floquet treatment is not required. However, in the neighborhood of $k\smeq 0$ the system does need a Floquet treatment. The energy of the solutions for $k\rightarrow 0$ are $\pm \sqrt{\Delta_0^2+ \mu^2}$. For a $\hbar\Omega = 2\sqrt{\Delta_0^2+ \mu^2}$, the energies of Floquet blocks with $(m\smmi m')\smeq 1$ cross at $k\smeq 0$. In particular, the $(m,-)$ band crosses the $(m\smmi 1,+)$ one. This situation is illustrated in Fig.\ref{FG:fig1}(C.a) as the upper band in solid black line ($m' \smeq 0$) crosses the lower band in dashed black line ($m\smeq 1$). From this condition we define the critical frequency, $\Omega_{\rm c}$, as
\be
\Omega_{\rm c}\equiv\frac{2}{\hbar}\sqrt{\Delta_0^2+\mu^2}.
\label{EQ:critFreq}
\ee

In Figs.~\ref{FG:fig1}(C.b) and \ref{FG:fig1}(C.c) we show the resulting Floquet quasienergies at the critical frequency for two different amplitudes of the rotating SOC. For small $k$ we obtain Dirac-cone like quasienergy excitations closing at values
\be
\ve_{0,n}(\Omega) \equiv(n+1/2)\hbar\Omega\,,
\label{EQ:halfDriving}
\ee
with $\Omega=\Omega_{\rm c}$. They coexist with other bands also closing at $\ve_{0,n}(\Omega_{\rm c})$, but with vanishing $\partial\ve/\partial k$. The bands do not longer close at $k\smap 0$ for either $\Omega<\Omega_{\rm c}$ [shown in Fig.\ref{FG:fig1}(T.b) and Fig.\ref{FG:fig1}(T.c)] or $\Omega>\Omega_{\rm c}$ [shown in Figs.~\ref{FG:fig1}(B.b) and \ref{FG:fig1}(B.c)]. In the following we show that the latter two cases are associated with different Floquet topological phases.  Notice that the full quasienergy spectrum is gapless as the solutions for larger $k$ take all possible quasienergy values.

Indeed, the mapping introduced in Sec.\ref{SC:Urt} is useful for writing the Floquet solutions of $\hat{H}(t)$ in terms of the solutions of $\hat{H}_{0}$ (the all-static 1D superconducting system with SOC and Zeeman interaction). We start from the eigenstates of $\hat{H}_{0}$ with energy $E_a$ [see Eqs.~(\ref{EQ:SOC}) and (\ref{EQ:H0})]
\be
 \ketLR{\phi^0_a(t)} \rightarrow {\rm e}^{-\frac{\ci}{\hbar}E_a t}  {\it\bar{\Phi}}^0_a(x),
\ee
where ${\it\bar{\Phi}}^0_a(x)$ is a Nambu spinor satisfying $(\hat{H}_{0}-E_a) {\it\bar{\Phi}}^0_a(x) \smeq0$. The $\hat{H}_{0}$ eigenenergies for the infinite case are distributed in four bands,\cite{SauDasSarma2010,*Alicea2010prb,*OregRefaelvonOppen2010}
\bea
E_a=E_{b_1,b_2}(k)&=&b_1 \left[\Delta_0^2+E_{\rm Z}^2 +\left(\frac{\hbar^2k^2}{2m^*}-\mu\right)^2 +\alpha^2 k^2\vphantom{\sqrt{{\xi_k^0}^2}} \right. \label{EQ:EnergiesH0}\\&& \left.+\vphantom{\sqrt{{\xi_k^0}^2}} b_2\sqrt{\Delta_0^2 E_{\rm Z}^2 + \left(\frac{\hbar^2k^2}{2m^*}-\mu\right)^2 \left(\alpha^2 k^2+E_{\rm Z}^2\right)}\right]^{1/2} \,,~~~~\nonumber
\eea
with $b_{1}\smeq\pm 1$ and $b_{2}\smeq\pm 1$. The spin rotation of Eq.\eqref{EQ:UR} applied on a generic static solution gives
\be
\ketLR{\phi_a(t)}\equiv \hat{U}_\mathrm{R}^\dagger(t)   \ketLR{\phi^0_a(t)}~\rightarrow~~  {\it\bar{\Phi}}_{a}(x,t),
\label{EQ:rotated}
\ee
which is a solution of the non-magnetic IR SOC superconducting 1D system by substituting $E_Z \rightarrow \hbar\Omega/2$ and $\alpha\rightarrow\alphatilde$.

The physical state $\ketLR{\phi_a(t)}$, Eq.\eqref{EQ:physical}, can be associated with a family of Floquet QESs. The members of such family have quasienergies differing in multiples of $\hbar\Omega$. Our goal is to determine the quasienergy values associated with solutions with energy $E_a$ of the static problem by using Eq.\eqref{EQ:rotated}. For convenience, we first write each eigenstate for the static system of Ref.~\onlinecite{SauDasSarma2010,*Alicea2010prb,*OregRefaelvonOppen2010} (assumed to be known) as a Nambu spinor, ${\it\bar{\Phi}}^0_a(x)=\left(A_a(x),B_a(x),C_a(x),D_a(x)\right)^T$. We cast the resulting physical state in the form $\ketLR{\phi_a(t)}= {\rm e}^{-\frac{\ci}{\hbar}\ve_a t} \ketLR{\phi^T_a(t)}$. This allows for the identification of the quasienergy for the time-dependent system,
\be
\ve_a = E_a + (n + 1/2)\hbar\Omega= E_a + \ve_{0,n}(\Omega) ,
\label{EQ:mappingQE}
\ee
and the associated Floquet QES as the Nambu spinor,
\be
{\it\bar{\Phi}}^T_a(x,t)={\rm e}^{\ci n \Omega t}\left(A_a(x){\mathrm{e}}^{\ci \Omega t },B_a(x),C_a(x){\mathrm{e}}^{\ci \Omega t },D_a(x)\right)^T\!.
\label{EQ:mappingQEstates}
\ee
Importantly, Eqs.\eqref{EQ:mappingQEstates} and \eqref{EQ:mappingQE} are valid also for cases in which both linked systems are not translational invariant and thus the solutions are not eigenstates of the momentum $p_x$ (related examples on finite and disordered systems are discussed in Sec.\ref{SC:finite}).

We set $n\smeq 0$ in Eq.\eqref{EQ:mappingQE} to generate Floquet quasienergies of the IR case using the energies of the static system given in
Eq.\eqref{EQ:EnergiesH0}.  We plot in black solid lines the resulting $\ve_a$ in panels (b) and (c) of Fig.\ref{FG:fig1}, i.e., for different driving frequencies and Rashba strengths.
Floquet quasienergies obtained entirely within the Floquet picture with our numerical method show an excellent agreement with the analytical results.
In Figs.~\ref{FG:fig1}(C.b) and \ref{FG:fig1}(C.c) for $\Omega=\Omega_{\rm c}$ we see that the Floquet-Dirac branches closing at $\ve_{0,n}(\Omega_{\rm c})$ map to the Dirac-cone branches of $\hat{H}_{0}$ that close at $E_a\smeq 0$ (those with nonvanishing $\partial \ve/\partial k$ at $k\smeq 0$). In the plot, we have chosen $\mu>\Delta_0$ to show that the mapping holds in general. It is known that the $\mu\smeq 0$ case is the most favorable for expressing the Kitaev model in the static $\hat{H}_{0}$ because the effective \emph{p}-wave gap is larger.\cite{AliceaNatPhys2011} Here, $\mu>\Delta_0$ assures a nonzero electron density allowing the development of superconductivity by proximity effect when switching off the
driving ($\alphatilde\smeq 0$).

In the static system, as discussed above, the closing and reopening of the band gap indicate a change in topological properties. There, when MFs are present, they appear as solutions of the BdG equation with $E_a\smeq E_{\rm MF}\smeq 0$. This follows from the particle-hole symmetry of the BdG Hamiltonian that enforces zero energy for any eigenstate which is its own antiparticle. In the translational invariant system $E_a\smeq0$ solutions only appear at the critical Zeeman energy for $k\smeq 0$. Either in extended or finite systems, the general form of a Majorana fermion in Nambu space is\cite{FlensbergMajoranaReview2012} ${\it\bar{\Phi}}^0_{\rm MF}(x,t)\smeq\left(A(x),B(x),B^*(x),-A^*(x)\right)^T$, and the associated fermionic operator reads (since $E_a\smeq 0$, both the Schroedinger and the Heisenberg pictures lead to the same operator, see Eq.\eqref{EQ:annihil})
\bea
\hat{\Phi}^0_{\rm MF}= \left.\hat{\Phi} ^0_{\rm MF}\right.^\dagger&=&\int dx \left[ A(x) \hat{\Psi}_\uparrow(x) +A^*(x)     \hat{\Psi}^\dagger_\uparrow(x) \right.\nonumber \\ &&\left. +B(x) \hat{\Psi}_\downarrow(x) +B^*(x)    \hat{\Psi}^\dagger_\downarrow(x) \right].
\eea

A natural question arising here is whether the existence of Majorana solutions in the static system ensures the existence of Majorana solutions in the IR system of Eq.\eqref{EQ:H0td}. To see this, we construct the solution of the driven IR system which is associated with a MF solution of $\hat{H}_{0}$. In the Nambu spinor representation this is just ${\it\bar{\Phi}}_{\rm FMF}(x,t)\smeq  \hat{U}_\mathrm{R}^\dagger(t)   {\it\bar{\Phi}}^0_{\rm MF}(x,t)$, obtaining
\be
{\it\bar{\Phi}}_{\rm FMF}(x,t)= \left( A(x) {\mathrm{e}}^{\frac{\ci}{2} \Omega t },B(x) {\mathrm{e}}^{-\frac{\ci}{2} \Omega t }, B^*(x) {\mathrm{e}}^{\frac{\ci}{2} \Omega t },-A^*(x) {\mathrm{e}}^{-\frac{\ci}{2} \Omega t }\right).\label{EQ:FMajoranaF}
\ee
We observe that this state satisfies the particle-antiparticle condition while it evolves periodically in time [this follows from Eq.\eqref{EQ:annihil} noting that $\hat{\Phi}_H(t)\smeq \hat{\mathcal{U}'}^\dagger(t,0) \hat{\Phi}_H(0)  \hat{\mathcal{U}'}(t,0))\smeq \hat{\Phi}_H(t)^\dagger$]: It is a Floquet Majorana fermion (FMF). Notice that two driving periods are needed for the time-dependent state to revisit the same instantaneous configuration. This is so because the $\hat{U}_\mathrm{R}(t)$ of Eq.~(\ref{EQ:UR}) consists of spin-1/2 operators with rotation angle $2\pi t/T$ (see Sec.\ref{SC:Heff}). For $t\smeq 2n T$ (rotation angle $4\pi n$), it reduces to the identity. This is consistent with the fact that a representation within Floquet theory of such a MF state--- $E_a \smeq 0$ in Eq.\eqref{EQ:mappingQE}--- must have a quasienergy
\be
\ve_{\rm FMF}=\ve_{0,n}(\Omega)\smeq(n + 1/2)\hbar\Omega,
\label{EQ:qenFMF}
\ee
meaning that FMF states acquire a phase of $\pi \smpl 2\pi n$ (i.e., a factor $-1$) after one driving period. At $\Omega\smeq\Omega_{\rm c}$, the latter coincide with the quasienergies at which the Floquet-band gaps (those with $k\smap 0$) close.

Similar FMFs with finite quasienergy were reported by Jiang \emph{et al.} in Ref.~\onlinecite{Cirac2011prl}, where they studied a BEC cold-atom quantum wire (with static SOC and magnetic field) subject to a periodic in time chemical potential. Their results demonstrate that finite-quasienergy FMFs are not restricted to our particular model. An important point brought forward by Jiang \emph{et al.} was that the BdG-Floquet operator preserves electron-hole symmetry. In his context, this means that for each Floquet state with quasienergy $+\ve_a$ created with the operator ${\hat{\Phi}_a}^\dagger$ (see Eq.\eqref{EQ:annihil}) there is another one at $-\ve_a$ which is created with the operator $\hat{\Phi}_a$. Since Floquet states with quasienergies differing in $n\hbar\Omega$ address to the same set of physical solutions, the Majorana condition (${\hat{\Phi}_a}^\dagger\smeq {\hat{\Phi}_a}$) can also be satisfied by states exactly at the nonzero quasienergies given in Eq.\eqref{EQ:qenFMF}. We return to the discussion of unpaired FMFs in Sec.\ref{SC:finite} as bound states appearing at the edges of finite-size topological systems.

\subsubsection{The IR driving: Topological properties}
\label{SC:Heff}

In Eq.\eqref{EQ:mappingQE} [Eq.\eqref{EQ:mappingQEstates}] we have established a link between the eigenenergies $E_a$ [eigenstates $\ketLR{\phi^0_a(t)}$] of the static quantum wire of Ref.~\onlinecite{SauDasSarma2010,*Alicea2010prb,*OregRefaelvonOppen2010}, modeled by $\hat{H}_{0}$, and the quasienergies $\ve_a$ [Floquet states $\ketLR{\phi^T_a(t)}$] of our time-dependent proposal, modeled by $\hat{H}(t)$. Similarly in Eq.\eqref{EQ:FMajoranaF} we have shown, by using an exact mapping, that for each Majorana solution appearing in the static system there is an associated FMF appearing in the driven system. Thus, the FMFs are protected by the \emph{same energy gap} as the MFs in the static system of Ref.~\onlinecite{SauDasSarma2010,*Alicea2010prb,*OregRefaelvonOppen2010}.

Here, in order to make explicit that the topological properties of the Floquet system are equivalent to those in the static system, we make use of our mapping to write the evolution operator for the Floquet problem, $\hat{\mathcal{U}}(t,0)$: The solutions can then be written as $\ketLR{\phi_a(t)}\smeq\hat{\mathcal{U}}(t,0) \ketLR{\phi_a(0)}$. First, because of the time-independence of $\hat{H}_{0}$, we write the time evolved state in the static system as
\be
 \ketLR{\phi^0_a(t)}= \exp{\left(-\frac{\ci}{\hbar} \hat{H}_{0} t \right)} \ketLR{\phi^0_a(0)} .
\label{EQ:evolStatic}
\ee
In order to simulate the time-dependent system subject to IR driving, the interaction strengths in Eq.\eqref{EQ:SOC} are to be replaced as $\alpha\rightarrow \tilde{\alpha}$ and $E_Z \rightarrow \hbar\Omega/2$. From Eq.\eqref{EQ:rotated} at $t\smeq 0$, we see that the mapping at $t\smeq 0$ is trivial, i.e., $\ketLR{\vphantom{\phi^0_a}\phi_a(0)}\smeq \ketLR{\phi^0_a(0)}$. Then, by introducing Eq.\eqref{EQ:evolStatic} in Eq.\eqref{EQ:rotated}, the evolution operator for the time-dependent non-magnetic system becomes
\be
\hat{\mathcal{U}}(t,0) \equiv \hat{U}^\dagger_\mathrm{R}(t)   \exp{\left(-\frac{\ci}{\hbar} \hat{H}_{0} t \right)}\,.
\label{EQ:rotated2}
\ee
After replacing $E_Z \rightarrow \hbar\Omega/2$ in Eq.\eqref{EQ:UR} we get $\hat{U}^\dagger_\mathrm{R}(T)  \smeq -\eins_4$. This corresponds to a $2\pi$ angle spin rotation, with $\eins_4$ the identity in the Nambu space (in the same basis as $\hat{H}_{0}$, given in Eq.\eqref{EQ:HBdG}). Therefore, when particularizing $\hat{\mathcal{U}}(t,0)$ for the evolution over one full period we obtain
\bese
\bea
\hat{\mathcal{U}}(T,0)&=& \exp{\left(-\frac{\ci}{\hbar}\hat{H}_{\rm eff} T \right)}, \\
\hat{H}_{\rm eff}&\equiv&\hat{H}_{0}+\eins_4 \ve_{0,n}(\Omega).
\eea
\label{EQ:Heff}
\eese
This means that the evolution over one period in the time-dependent system, $\hat{\mathcal{U}}\left(T,0\right)\smeq {\bm{T}}_t\exp{(-\frac{\ci}{\hbar} \! \int_{0}^T  \hat{H}(t') dt' )}$, is equivalent to the evolution over the same period $T$ with the time-independent effective Hamiltonian, $\hat{H}_{\rm eff}$. The effective Hamiltonian is just $\hat{H}_{0}$ trivially shifted in energy by the amount $\ve_{0,n}(\Omega) \smeq (n\smpl 1/2)\hbar\Omega$  defined in Eq.\eqref{EQ:halfDriving}. In the light of Eq.\eqref{EQ:evolOp}, such shift justifies the form of the quasienergies in Eq.\eqref{EQ:mappingQE}.

As pointed out by Kitagawa \emph{et al.} in Ref.~\onlinecite{Kitagawa2010prb}, the topological properties contained in the effective Hamiltonian of a Floquet system are to be studied with the tools for time-independent topological phase transitions. The topological phases appearing in $\hat{H}_{\rm eff}$ imply the presence associated topological phases in the Floquet system. We can now formally conclude that the $Z_2$ topological invariant\cite{Kitaev2001} that distinguishes the two gapped phases in the magnetostatic model of Ref.~\onlinecite{SauDasSarma2010,*Alicea2010prb,*OregRefaelvonOppen2010} also distinguishes the two Floquet topological phases of the non-magnetic driven system proposed here. In particular, this corroborates that the regime with $\Omega>\Omega_{\rm c}$ (supercritical) is indeed a Floquet TSP whereas $\Omega<\Omega_{\rm c}$ (subcritical) is to be associated with the topologically trivial superconducting phase.

\subsubsection{The IR driving: Mean energies}
\label{SC:IRmean}

We now discuss the mean energies of the Floquet states depicted in Figs.~\ref{FG:fig1}(T.d), \ref{FG:fig1}(C.d) and \ref{FG:fig1}(B.d). These follow from Eq.\eqref{EQ:meanEnergy} after calculating the derivatives $\partial \varepsilon_a/\partial\Omega$. The numerical results agree with the analytical mean energies derived from Eq.\eqref{EQ:EnergiesH0} by computing
\be
 \overline{E}_a = E_a - E_Z \frac{\partial E_a}{\partial E_Z}\,,
 \label{EQ:MeanEnergyStatic}
\ee
and substituting $E_Z\rightarrow \hbar\Omega/2$. At critical frequency, $\Omega\smeq\Omega_{\rm c}$,
the two Floquet states with quasienergy forming a Dirac cone--- see black solid lines in Figs.~\ref{FG:fig1}(C.b) and \ref{FG:fig1}(C.c)--- lead to linear mean-energy dispersions around $k=0$ as shown in Fig.~\ref{FG:fig1}(C.d). We verified that this happens as a general rule, where the mean-energy bands closing the gap at $\overline{E}=0$ derive from the
Floquet-quasienergy bands closing at $\ve_{0,n}(\Omega_{\rm c})$.

\begin{figure}[!t]\begin{center}
\includegraphics[clip,width=0.35\textwidth]{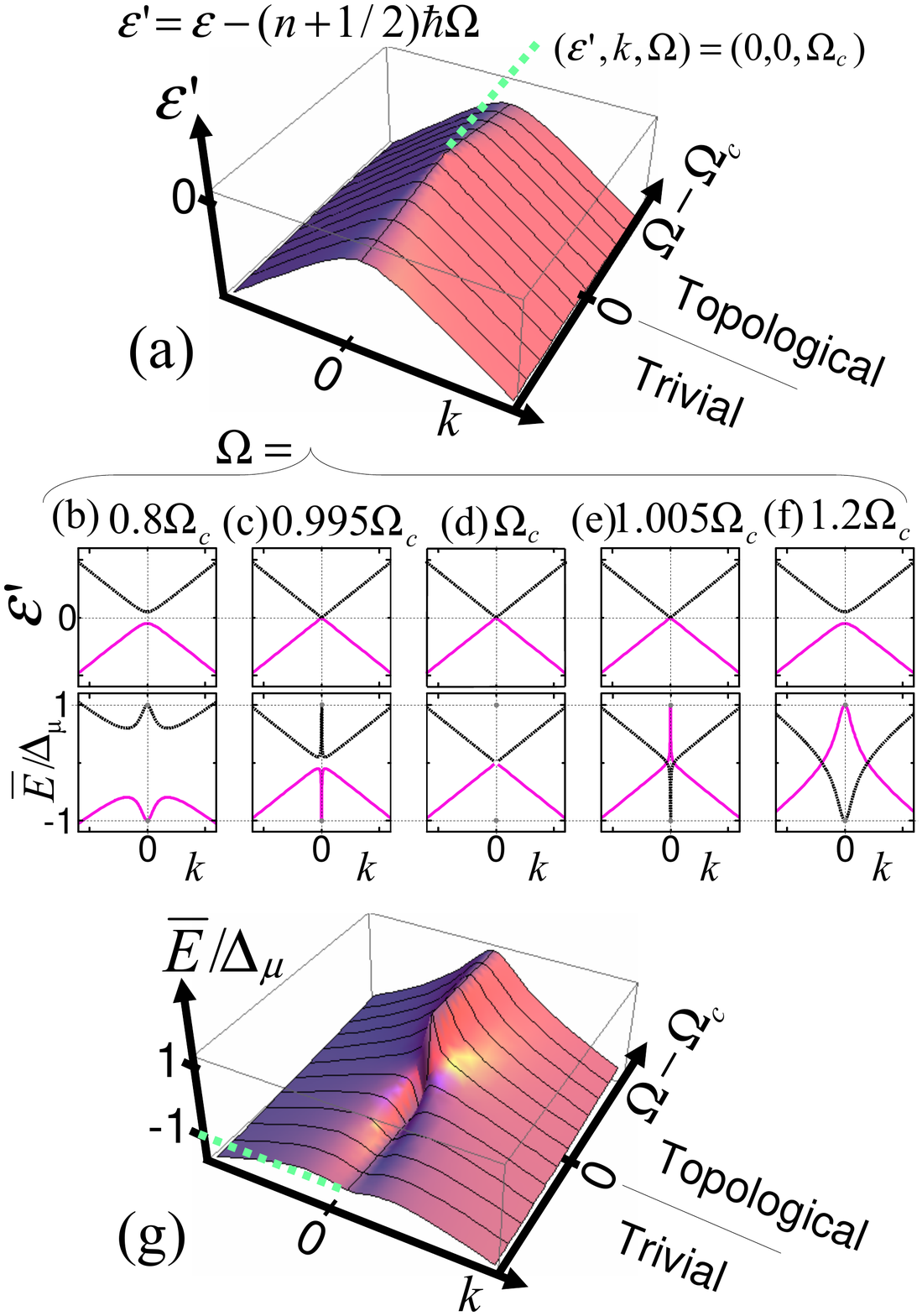}
\caption{\label{FG:fig2} (a) Quasienergies $\ve'\smeq\ve-\ve_{0,n}(\Omega)$ and (g) mean energies $\overline{E}$ of Floquet states as a function on the driving frequency $\Omega$ and the linear momentum $\hbar k$. In the plot $\Delta_\mu\smequiv\hbar\Omega_{\rm c}/2\smeq\sqrt{\Delta_0^2\smpl\mu^2}$. We focus on the solutions that are affected by the ideal rotating SOC driving in the small $k$ limit for  $\mu\ll\Delta_0$, given in Eq.\eqref{EQ:QESaffected}. For the sake of clarity, surface plots show the lower energy bands only (see $\ve_3(k)$ and $\overline{E}_3(k)$ in the text): Upper bands follow by symmetry ($\ve_4(k)\smeq-\ve_3(k)$ and $\overline{E}_4(k)\smeq-\overline{E}_3(k)$).
In panels (b), (c), (d), (e) and (f), the driving frequency $\Omega$ is fixed and both bands are shown. At the critical frequency--- panel (d)--- the quasienergies form a Dirac-cone while the mean energies show a similar behavior except for a discontinuity at $k\smeq0$.
Just away from $\Omega_{\rm c}$, sub- and super-critical regimes can not be distinguished from the quasienergy bands alone. In contrast, thanks to the discontinuity developed at the critical point, the mean-energy bands show remarkable features distinguishing one regime from another. Such a contrast is clearly seen already in the surface plots of panels (a) and (g).
}
\end{center}
\vspace{-0.4cm}
\end{figure}

As a result of a discontinuity in the mean-energy bands as a function of the driving frequency, these get very sensitive at $\Omega_{\rm c}$ for small $k$. Such a discontinuity is intimately related to the change of topological properties. Right at $k\smeq 0$, the amplitude of periodic driving vanishes and the mean energy simply reduces to the energy, $\overline{E}_\pm(k \smeq 0)\smeq \pm \sqrt{\Delta_0^2+ \mu^2}\smeq\hbar\Omega_{\rm c}/2$.
Below the critical frequency, the top mean-energy band is connected to the $\overline{E}_+(k \smeq 0)$ point while the bottom band acts likewise with the $\overline{E}_-(k \smeq 0)$ point, instead. Remarkably, above the critical frequency this relationship is inverted. The band inversion happens for  $\Omega\smeq\Omega_{\rm c}$, where linear dispersion in mean energy is developed excluding the states with $k=0$: Points $\overline{E}_\pm(k \smeq 0)$ appear isolated from the bands (see following discussion and Fig.~\ref{FG:fig2} for details). As a rule, the mean-energy spectrum is complicated for higher $k$. For $\mu\smap 0$, the mean-energy bands resemble those for the energy in the static system around the critical condition $\Omega\smeq\Omega_{\rm c}$. Away from this situation, the low $k$ mean energies are always different from the energies of the static system.\footnote{From Eq.\eqref{EQ:MeanEnergyStatic} the difference between the mean energy of a Floquet state and the energy of the associated eigenstate in the static system is directly proportional to the expectation value of $\sigma_z$ in the latter eigenstate.}

For a better understanding of criticality in the vicinity of $\Omega_{\rm c}$, it is useful to solve the Eq.\eqref{EQ:floqMatrix} for the Floquet operator in the limits of $\mu\ll\Delta_0$ and small $k$.
We first need the solutions of the static component of $\hat{H}(t)$, i.e., $\hat{H}^{(0)}$, in the Nambu basis with spin quantized along the 3-axis, $(\psi^\dagger_{\uparrow},\psi^\dagger_{\downarrow},\psi_{\downarrow},-\psi_{\uparrow} )^\dagger$. For simplicity, we define Nambu unit vectors as $\widehat{\psi}_l\smequiv (\delta_{1,l},\delta_{2,l},\delta_{3,l},\delta_{4,l})^T$. $\hat{H}^{(0)}$ only includes the kinetic term and the pairing potential; by defining the angle $\varphi_{k}\smequiv \arctan\left(\Delta_0/(\hbar^2 k^2/(2 m^*)-\mu)\right)$ we write $\hat{H}^{(0)}$ solutions as
\bese
\bea
\ketLR{+, k \uparrow} &~\rightarrow~~&\frac{{\rm e}^{\ci k x}}{\sqrt{L}}\left( \cos{\frac{\varphi_{k}}{2}} \widehat{\psi}_1 + \sin{\frac{\varphi_{k}}{2}} \widehat{\psi}_3 \right),
\\ \ketLR{+, k \downarrow} &~\rightarrow~~&\frac{{\rm e}^{\ci k x}}{\sqrt{L}}\left( \cos{\frac{\varphi_{k}}{2}} \widehat{\psi}_2 + \sin{\frac{\varphi_{k}}{2}} \widehat{\psi}_4 \right),
\\\ketLR{-, k \uparrow} &~\rightarrow~~&\frac{{\rm e}^{\ci k x}}{\sqrt{L}}\left( \sin{\frac{\varphi_{k}}{2}} \widehat{\psi}_1 - \cos{\frac{\varphi_{k}}{2}} \widehat{\psi}_3 \right),
\\ \ketLR{-, k \downarrow} &~\rightarrow~~&\frac{{\rm e}^{\ci k x}}{\sqrt{L}}\left( \sin{\frac{\varphi_{k}}{2}} \widehat{\psi}_2 - \cos{\frac{\varphi_{k}}{2}} \widehat{\psi}_4 \right),
\eea
\label{EQ:eigenstates1}
\eese
with the energies given in Eq.\eqref{EQ:NakedEnergies}. These states are, indeed, eigenstates of each block $\hat{H}^{(0)}\smpl m \hbar\Omega$ of the Floquet operator given in Eq.\eqref{EQ:floqMatrix}.  We define $\ketLR{m,\pm, k \sigma} \equiv {\rm e}^{\ci m \Omega t}\ketLR{\pm, k \sigma}$ for referring to the eigenstates of block $m$ by simply introducing a Fourier phase factor.

We limit our analysis only to blocks $m\smeq 0$ and $m\smeq 1$ in Eq.\eqref{EQ:floqMatrix}, assuming the mixing of few bands only. This is justified since: (i) The only mixing components in $\hat{H}(t)$ due to the periodic driving are $\hat{H}^{(\pm 1)}$, (ii) in the small-$k$ limit the time-dependent components are small, discouraging mixing of states far apart in energy, (iii) close to the critical frequency, $\Omega\smeq\Omega_{\rm c}\smpl \delta \epsilon/\hbar$, other states with small $k$ are at least $\hbar\Omega_{\rm c}$ away in quasienergy, and (iv) other choices with $m\smmi m'\smeq 1$ lead to shifted quasienergies (in multiples of $\hbar\Omega$) and, therefore, to equivalent solutions. Furthermore, we restrict the Floquet operator to the subspace generated by $\ket{0,+,k \sigma}$ and $\ket{1,-,k \sigma}$, i.e., the four states crossing when $\Omega\smeq \Omega_{\rm c}$.

Combining Eqs.~\eqref{EQ:HFourier1} and \eqref{EQ:HBdGfulltd} up to linear order in $k$, the Floquet operator of Eq.\eqref{EQ:floqMatrix} reduces to
\be
\widetilde{H}_{\rm F} =\left(\begin{array}{cccc}
\hbar\Omega_{\rm c}/2&0&0&0\\
0&\hbar\Omega_{\rm c}/2&\alphatilde k  \sin{\varphi_0}&0\\
0&\alphatilde k  \sin{\varphi_0}&\hbar\Omega_{\rm c}/2+\delta\epsilon&0\\
0&0&0&\hbar\Omega_{\rm c}/2+\delta\epsilon
\end{array} \right),
\label{EQ:HFrestrictedAlpha0}
\ee
in the ordered basis $\{\ket{0,+,k \uparrow}, \ket{0,+,k \downarrow}, \ket{1,-,k \uparrow}, \ket{1,-,k \downarrow}\}$. We readily notice that the 1st and 4th states
\bese
\bea
\ket{\phi^T_{1}(t)} &=& \ket{+,k \uparrow}  \mathrm{~~~~~~~~with~~}\varepsilon_{1}=\hbar\Omega_{\rm c}/2, \\
\ket{\phi^T_{2}(t)} &=& \ket{-,k \downarrow} {\rm e}^{\ci \Omega  t} \mathrm{~~~~with~~}\varepsilon_{2}=\hbar\Omega_\mathrm{c}/2 +\delta\epsilon,
\eea
\eese
do not mix under the action of the rotating SOC axis.
Their mean energy, Eq.\eqref{EQ:meanEnergy}, is identical to their energy:
\be
\overline{E}_1= \hbar\Omega_{\rm c}/2~,~~\overline{E}_2= -\hbar\Omega_{\rm c}/2,
\ee
where we used that $\partial \Omega_{\rm c}/ \partial\Omega=0$ and $\partial (\delta\epsilon)/\partial\Omega=\hbar$. The remaining two solutions, for $k\smneq 0$, are a mixture of states with different $m$. By defining the angle $\eta_k \smequiv \arctan\left(2\alphatilde k  \sin{\varphi_0}/\delta\epsilon\right)$, these solutions read
\bese
\bea
\ket{\phi^T_{3}(t)} &=& \cos{\frac{\eta_k}{2}}\ket{+,k \downarrow} - \sin{\frac{\eta_k}{2}} \ket{-,k \uparrow} {\rm e}^{\ci \Omega  t}, \\
\ket{\phi^T_{4}(t)} &=& \sin{\frac{\eta_k}{2}}\ket{+,k \downarrow} + \cos{\frac{\eta_k}{2}} \ket{-,k \uparrow} {\rm e}^{\ci \Omega  t},~~~~~~
\eea
\label{EQ:QESaffected}
\eese
with quasienergies
\be
\varepsilon_{3,4}(k)=\hbar\Omega_{\rm c}/2 + \frac{1}{2} \delta\epsilon\mp\sqrt{\left(\frac{\delta\epsilon}{2}\right)^2 + \left(\alphatilde k  \sin{\varphi_0}\right)^2}\!.
\ee
At the critical frequency, $\delta\epsilon\smeq 0$, the two branches form a Dirac cone in quasienergy. The vertex at $k\smeq0$ lies at quasienergy $\hbar\Omega_{\rm c}/2$ (while the $l$-shifted solutions cross at $\hbar\Omega_{\rm c}/2 \smpl l \hbar \Omega_{\rm c}\smeq\varepsilon_{l,0}(\Omega_{\rm c})$). The superconducting gap is essential since the mixing term $\alphatilde k  \sin{\varphi_0}$, with $\sin{\varphi_0} \smeq  -\Delta_0 / \sqrt{\Delta_0^2+\mu^2}$, vanishes for $\Delta_0\smeq 0$ (leading to flat dispersions).

For arbitrary values of $k$ (always in the small $k$ approximation) the mean energies for these two solutions are
\be
\overline{E}_3(k)=-\overline{E}_4(k)=\frac{\delta\epsilon\hbar\Omega_{\rm c}/2  -2 \left(\alphatilde k  \sin{\varphi_0}\right)^2}{\sqrt{\delta\epsilon^2+4 \left(\alphatilde k  \sin{\varphi_0}\right)^2 }}.
\ee
At the critical frequency,  $\delta\epsilon\smeq 0$, the two branches form a Dirac cone in mean energy, $\pm |\alphatilde k  \sin{\varphi_0}|$. This excludes the case $k\smeq0$, with mean energies fixed at $\pm\hbar\Omega_{\rm c}/2$. We find
\be
\overline{E}_3(0)=-\overline{E}_4(0)=\begin{cases}
-\frac{\hbar}{2}\Omega_{\rm c}, &\text{\!\!\!for }\Omega<\Omega_{\rm c}\\
\frac{\hbar}{2}\Omega_{\rm c}, & \text{\!\!\!for }\Omega>\Omega_{\rm c}.
\end{cases}
\ee
In Fig.\ref{FG:fig2} we show the different behavior of the quasienergies and mean energies involved around the transition, i.e., the solutions $3$ and $4$ in the latter derivation. For simplicity, only the bottom quasienergy [mean-energy] band is plotted on the $(\Omega,k)$ parameter space in Fig.\ref{FG:fig2}(a) [Fig.\ref{FG:fig2}(g)]. The low-$k$ quasienergy band passes through the transition without visible change (see upper panels in Fig.\ref{FG:fig2}(b) to Fig.\ref{FG:fig2}(f)). In contrast, the mean-energy band (lower panels in Fig.\ref{FG:fig2}(b) to Fig.\ref{FG:fig2}(f)) presents a clear distinction between the two sides of the transition. Both the abrupt change in mean energy and the Dirac-cone dispersion at the critical parameter are due to the band inversion and linear-in-$k$ mixing underlying the phase transition. This result is consistent with what is commonly found in Floquet systems, i.e., avoided crossings in quasienergy producing abrupt changes in the associated mean energies.\cite{FloquetFaisal1997}

In our study of the Floquet problem, the need to compute the mean energies $\overline{E}_a$ arises naturally. The emergence of the mean energy as an indicator of the band inversion is remarkable. It is therefore relevant to ask ourselves about the meaning of $\overline{E}_a$ in static systems as those discussed in Ref.~\onlinecite{SauDasSarma2010,*Alicea2010prb,*OregRefaelvonOppen2010} or even in more general contexts. We have shown that $\overline{E}_a$ can be derived from Eq.\eqref{EQ:MeanEnergyStatic} as a function of the eigenvalues of the static system. By choosing an eigenstate $\ketLR{\phi^0_a(t)}$ of $\hat{H}_{0}$ and applying the Hellmann-Feynman theorem, we find that $\overline{E}_a$ is just the expectation value of the Hamiltonian on state $\ketLR{\phi^0_a(t)}$ provided the term proportional to $E_Z$ is \emph{missing}, i.e., disregarding the Zeeman coupling. In other words, $\overline{E}_a$ is the expectation value on $\ketLR{\phi^0_a(t)}$ (eigenstate of the full Hamiltonian) of the operator obtained by enforcing $E_{\rm Z}=0$ in $\hat{H}_{0}$. Similarly, for each energy eigenstate $\ketLR{\phi^0_a(t)}$, two other indicators of the band inversion can be constructed here: The expectation value of $\hat{H}_{0}$ evaluated at $\Delta_0=0$,
\be
I_{\Delta_0} =E_a - \Delta_0 \frac{\partial E_a}{\partial \Delta_0},
 \label{EQ:MeanEnergyStaticDelta}
\ee
and the expectation value of $\hat{H}_{0}$ evaluated at $\mu=0$,
\be
I_{\mu} = E_a - \mu \frac{\partial E_a}{\partial \mu}.
 \label{EQ:MeanEnergyStaticMu}
\ee
We verified that the bands generated by these expectation-values, $I_\zeta$ (with $\zeta\smeq \{\mu,E_Z,\Delta_0\}$ and $I_{E_Z}\smeq\overline{E}_a$), change abruptly at the topological phase transition. Physically, this is not surprising since the complements of $I_\zeta$, namely, $E_a -I_\zeta$ are the expectation values on the energy eigenstates of the different contributing terms in the Hamiltonian. As discussed in Sec.\ref{SC:deriving}, the $k\smap 0$ eigenstates of the bands involved in the topological phase transition change abruptly when passing through the critical condition. This justifies the observed behavior. We do not comment any further on these indicators as that is beyond the scope of this work.

\subsubsection{Other forms of SOC driving beyond IR case}

\begin{figure}[b]\begin{center}
\includegraphics[clip,width=0.40\textwidth]{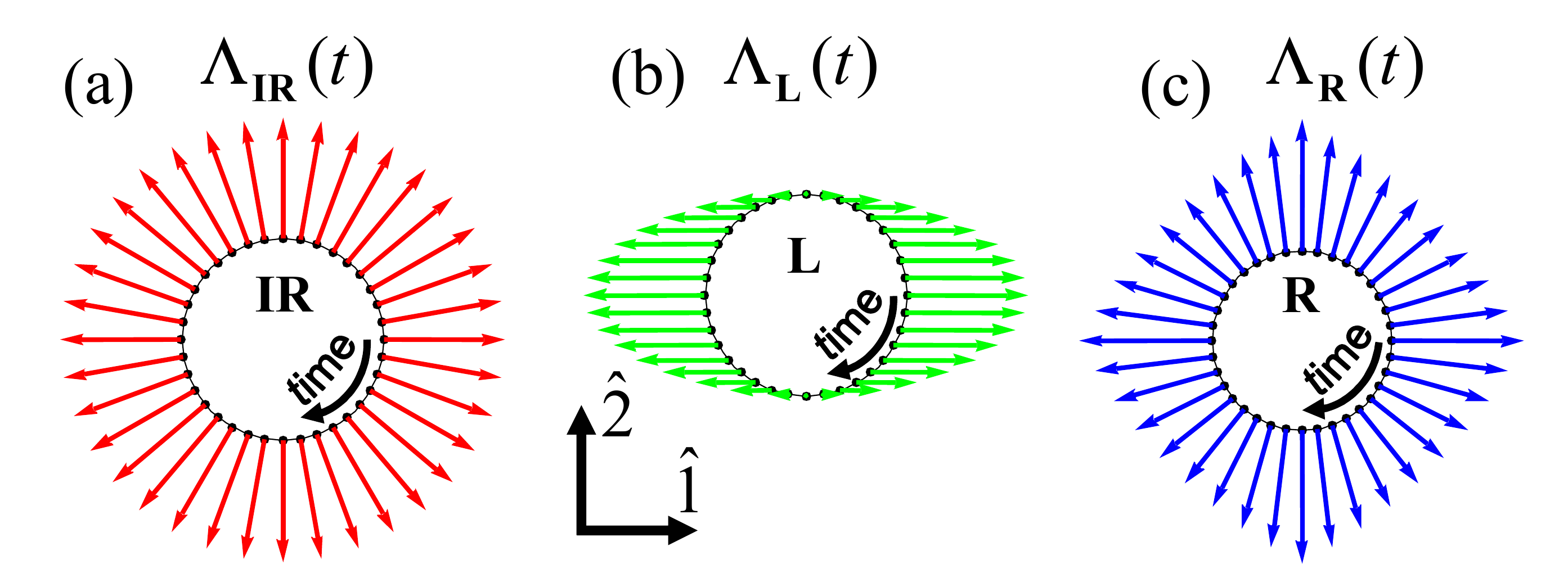}
\caption{\label{FG:fig3} Sketch of three different cases for the periodic SOC driving $\mathbf{\Lambda}(t)$ of Eq.~(\ref{EQ:HeHhTime}). (a) Ideally rotating (IR) driving $\mathbf{\Lambda}_\mathrm{IR}(t)$ of Eq.~(\ref{EQ:ideal}): The SOC axis rotates uniformly in the $1-2$ plane with constant coupling strength. (b) Linear (L) driving $\mathbf{\Lambda}_\mathrm{L}(t)$ of Eq.~(\ref{EQ:typeL}): The orientation of the SOC is fixed while the coupling strength oscillates harmonically. (c) Ramp (R) rotation driving $\mathbf{\Lambda}_\mathrm{R}(t)$ of Eq.~(\ref{EQ:typeR}): The SOC axis rotates anharmonically in the $1-2$ plane with varying coupling strength driven by symmetrical triangular waves}.
\end{center}
\end{figure}

We have shown that the topological system of Ref.~\onlinecite{SauDasSarma2010,*Alicea2010prb,*OregRefaelvonOppen2010} is directly related to a Floquet system composed of a non-magnetic superconducting wire subject to a rotating SOC axis. The studied time-dependent Hamiltonian satisfies: (i) Fourier decomposition has only zero frequency and $\pm\Omega$ components, (ii) the zero frequency component of the Hamiltonian has no SOC contribution, (iii) the ideal SOC rotation introduces a spin symmetry around the 3-axis, and (iv) the choice of one out of two possible rotational senses for the SOC axis determines the breaking of time-reversal symmetry (needed for emergence of a single MF at each edge of the topological region).\cite{FlensbergMajoranaReview2012} Here, we briefly discuss the Floquet quasienergies and mean energies in systems satisfying only some of the above conditions. Notice, however, that a more rigorous study of the situations below would require a detailed analysis of the effective Hamiltonian on each case (see Eq.\eqref{EQ:Heff}).

\begin{figure*}[!t]
\begin{center}
\includegraphics[clip,width=0.99\textwidth]{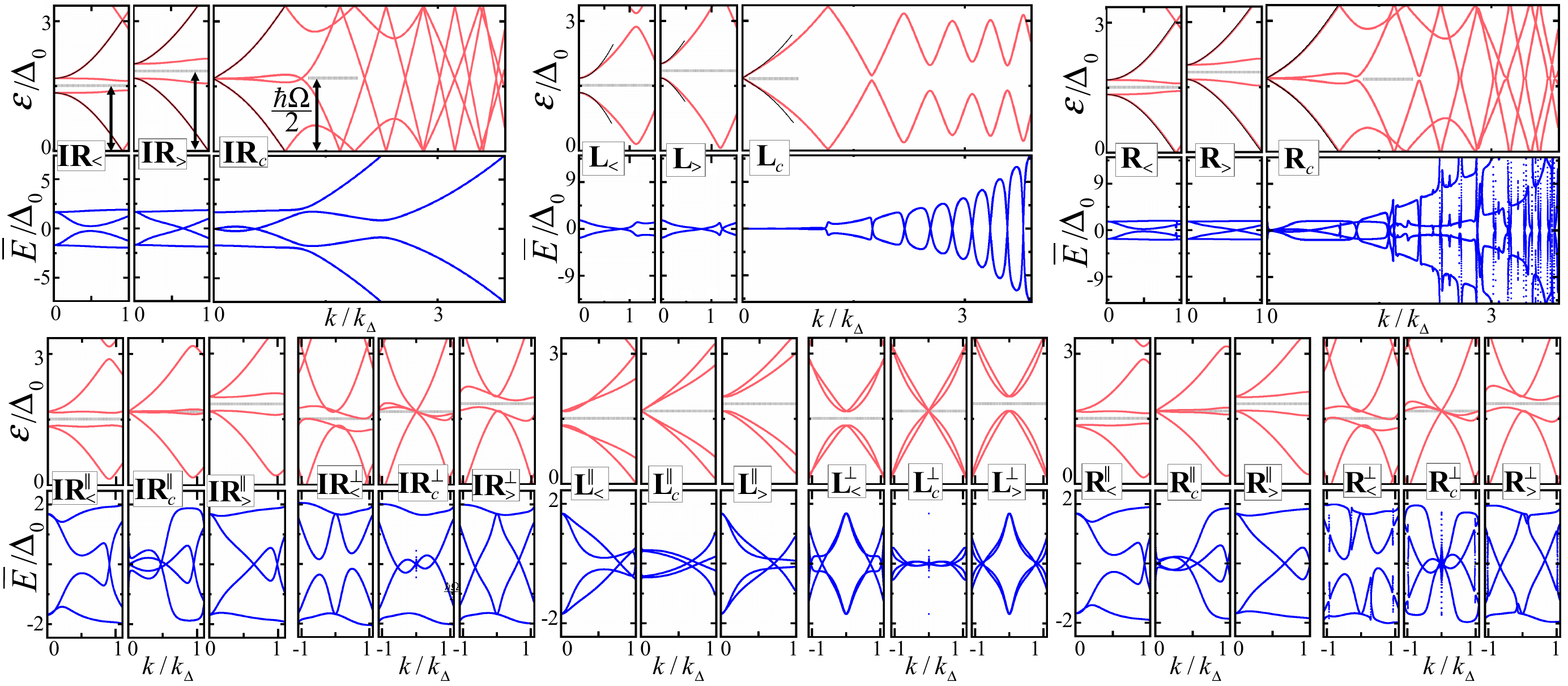}
\vspace{-0.4cm}
\caption{\label{FG:fig4} Floquet quasienergies $\ve(k)$ and mean energies $\overline{E}(k)$ for the situations listed in Table~\ref{TB:SOCcases} with $\mu\smeq 1.35\Delta_0$, $\hbar\Omega_\mathrm{c}\smap 3.36\Delta_0$, and $\tilde{E}_\mathrm{so}\smeq\alphatilde^2 m^*/(2 \hbar^2)\smeq 0.75\Delta_0$. Whenever present (lower panels), the static SOC component is set $\alpha_0\smeq\alphatilde/5$.
Labels subindices ``$<$", ``${\rm c}$" and ``$>$" stand for $\Omega =0.9 \Omega_{\rm c}$, $\Omega_{\rm c}$, and $1.1\Omega_{\rm c}$, respectively. Floquet quasienergies are shown from $0$ to $\hbar\Omega_{\rm c}$. Floquet bands are symmetrical with respect to $\ve\smeq  \hbar\Omega/2$ (horizontal grey line). We do not show the $-k$ quasienergies and mean energies whenever they are identical to the $+k$ ones. This is not the case for finite static SOC orthogonal to the periodic SOC driving. For vanishing static SOC (upper panels), solid black lines depict IR-quasienergy Floquet solutions--- in particular, the bands associated with the topological transition [quasienergies from Eq.\eqref{EQ:mappingQE} with $n\smeq 0$ generated by the energies of the static problem $E_{+,-}(k)$ and $E_{-,+}(k)$ given in Eq.\eqref{EQ:EnergiesH0}]. For L driving, the solutions with $\alpha_0\smeq0$ are doubly degenerated. For R driving, higher-frequency driving components generate new transitions. Floquet transitions introduce avoided crossings in the quasienergy spectrum and strong discontinuities in the mean energies. \cite{FloquetFaisal1997,FloquetHsu2006} See text for further discussion.}
\end{center}
\vspace{-0.6cm}
\end{figure*}

In Fig.~\ref{FG:fig3} we present a sketch of the ideal rotating SOC, panel (a), together with two alternative types of SOC drivings, panels (b) and (c). Figure \ref{FG:fig3}(b) depicts a linear (L) driving $\mathbf{\Lambda}_\mathrm{L}(t)$, such that
\bea
\mathbf{\Lambda}_\mathrm{L}(t) \cdot\boldsymbol{\sigma}
&\equiv&
\mathbf{\Lambda}_\mathrm{IR}(t)\cdot\boldsymbol{\sigma}+\sigma_1(\mathbf{\Lambda}_\mathrm{IR}(t)\cdot\boldsymbol{\sigma})\sigma_1, \nonumber\\
\Rightarrow~~~ \mathbf{\Lambda}_\mathrm{L}(t)&=&\left(2\alphatilde\cos \left(\Omega t\right),0,0\right).
\label{EQ:typeL}
\eea
This represents the superposition of clockwise and counterclockwise IR rotations with the same amplitude $\alphatilde$. This driving does not fulfill property (iii) of the IR case since the SOC strength oscillates while the axis direction is preserved. The Fourier amplitudes are
\be
\hat{H}^{(\pm 1)}_\mathrm{e,L}=\hat{H}^{(\pm 1)}_\mathrm{h,L} = 2\frac{\alphatilde}{\hbar} p_x \sigma_1 \smeq 2\frac{\alphatilde}{\hbar} p_x \left(\sigma_++\sigma_-\right). \label{EQ:HFourier2}
\ee
This means that the reduced Floquet operator for the L driving mixes not only states $\ket{0,+,k \downarrow}$ and $\ket{1,-,k \uparrow}$, as in the IR case of Eq.\eqref{EQ:HFrestrictedAlpha0}, but also states $\ket{0,+,k \uparrow}$ and $\ket{1,-,k \downarrow}$. This extra mixing is a consequence of the violation of property (iv), leading to the double-degeneracy of the full quasienergy spectrum, as we see below.

Figure \ref{FG:fig3}(c) depicts a different kind of rotation defined by
\be
\mathbf{\Lambda}_\mathrm{R}(t)\equiv \left(\alphatilde \mathrm{cos_R} \left(\Omega t\right),-\alphatilde \mathrm{sin_R} \left(\Omega t\right),0\right),
\label{EQ:typeR}
\ee
which is based on a $2\pi$-periodic ramp (R) or symmetrical triangular wave function
\bese
\bea
\mathrm{sin_R}(x)&\equiv& \begin{cases}
 \frac{\pi (-1)^{m}}{4} \left(x-m\pi \right), &\text{\!\!\!for }-\frac{\pi}{2} \leq x-m\pi<\frac{\pi}{2}~ \end{cases}\nonumber \\
&=& \sum_{n>0,\text{odd}} \frac{(-1)^{\frac{n-1}{2}}}{n^2}\sin(n x), \\
\mathrm{cos_R}(x)&\equiv& \mathrm{sin_R}(x+\pi/2).
\eea
\eese
With this choice--- notice that $\mathrm{cos_R}(n\pi)\smeq(-1)^n\pi^2/8$---, the contributions at frequencies $\pm\Omega$ are identical to those for IR case with Fourier amplitudes $\hat{H}^{(\pm 1)}_\mathrm{e/h,R}\smeq\hat{H}^{(\pm 1)}_\mathrm{e/h}$ given in Eq.~(\ref{EQ:HFourier1}). Higher frequency Fourier components are
\be
\hat{H}^{(\mp n)}_\mathrm{e,R}=\hat{H}^{(\mp n)}_\mathrm{h,R} = \frac{(-1)^{\frac{n-1}{2}}}{n^2}\frac{\alphatilde}{\hbar} p_x \sigma_\pm,
\label{EQ:HFourier3}
\ee
for $n$ odd. The R driving, hence, violates properties (i) and (iii) of the IR case.

So far we have considered three different types of driving $\mathbf{\Lambda}(t)$, without any static SOC component. We now allow a nonzero SOC component of strength $\alpha_0$ through the vector $\mathbf{\Lambda}^{(0)}\smeq \hat{n}_0 \alpha_0/\hbar$ in Eq.\eqref{EQ:HeHhTime}. The unit vector $\hat{n}_0$ is assumed either parallel or perpendicular to the plane defined by $\mathbf{\Lambda}(t)$. In the following, we focus on the case of dominating $\mathbf{\Lambda}(t)$ such that $\alpha_0 \sim 0.2 \alphatilde$ (nothing particular is expected for $\alpha_0 \gg \alphatilde$ since the system behaves, essentially, as a non-topological wire with static SOC and vanishing magnetic field). In Table \ref{TB:SOCcases} we define the labels used to identify $9$ different situations addressed numerically with the method discussed in the Appendix \ref{AP:A}. Notice that drivings L and R, together with most cases with nonvanishing $\mathbf{\Lambda}^{(0)}$, can not be easily mapped to a time-independent model.

\begin{table}[!t]
\caption{Labels used to identify different SOC drivings $\mathbf{\Lambda}(t)$ in the presence of a static SOC component $\mathbf{\Lambda}^{(0)}$.}
\begin{tabular}{l||c|c|c}
\hline
\hline
$\mathbf{\Lambda}^{(0)}$~(below),~$\mathbf{\Lambda}(t)~\rightarrow~$
&$\mathbf{\Lambda}_\mathrm{IR}(t)$ &$\mathbf{\Lambda}_\mathrm{R}(t)$&$\mathbf{\Lambda}_\mathrm{L}(t)$\\
\hline
$\mathbf{\Lambda}^{(0)}=\mathbf{0}$&$\mathrm{IR}$ & $\mathrm{R}$&$\mathrm{L}$\\
$\mathbf{\Lambda}^{(0)}\parallel\mathbf{\Lambda}(t)$&$\mathrm{IR}^\parallel$& $\mathrm{R}^\parallel$ & $\mathrm{L}^\parallel$\\
$\mathbf{\Lambda}^{(0)}\perp\mathbf{\Lambda}(t)$&$\mathrm{IR}^\perp$& $\mathrm{R}^\perp$ & $\mathrm{L}^\perp$\\
\hline
\hline
\end{tabular}
\vspace{0cm}
\label{TB:SOCcases}
\end{table}

The results are presented in Fig.~\ref{FG:fig4}. At the critical point $\Omega\smeq\Omega_{\rm c}$, the Floquet-band gaps close at $\ve\smeq \ve_{0,0}(\Omega_{\rm c})\smeq \hbar \Omega_{\rm c}/2$ for $k \rightarrow 0$.
For L driving, as in the IR case, the Floquet-band gap (at small $k$) closes for the critical frequency forming Dirac-like cones. However, the bands are doubly degenerated and thus protected unpaired edge states are not possible in finite samples; on the other hand double degenerated localized states are not forbidden. However, here even in absence of local perturbations, localized states do not appear at the edges (see Sec.\ref{SC:finite}). For completeness, in Fig.\ref{FG:fig4} we also present results for the L driving coexisting with a static SOC where the referred double degeneracy is lifted. However, we have checked that no edge states can be found in the associated finite system (Sec.\ref{SC:finite}). This is something expected as we find two Floquet bands with finite slope closing the gap at the critical frequency.

For R driving, in the vicinity of $k\smeq 0$ the Floquet bands are similar to those for the IR case. For larger $k$, the effect of higher harmonics modifies significantly the shape of quasienergy bands. The avoided crossings in quasienergy produce abrupt changes in the mean energies, and the spikes in mean energies grow with the driving amplitude.\cite{FloquetFaisal1997,FloquetHsu2006} In Sec.~\ref{SC:finite} we show that related finite systems can hosts Majorana solutions at the edges, but presenting some signatures of the higher harmonics.
Considering that the amplitudes of the R-driving Fourier components (Eq.\eqref{EQ:HFourier3}) decay very fast, these results suggest that Majorana fermions at the edges might be absent for arbitrary SOC drivings with significant contributions from higher harmonics.

Finally, we briefly discuss the effect of a static SOC component with results shown in Fig.~\ref{FG:fig4}. For both
IR and R driving, a $\mathbf{\Lambda}^{(0)}$ orthogonal to the plane defined by the driven SOC axis produces a breaking of $\pm k$ symmetry.
Despite symmetry breaking, the main features discussed for vanishing $\mathbf{\Lambda}^{(0)}$ are maintained.
This is related to the fact that the IR-mapping $\hat{U}_\mathrm{R}^\dagger(t)$ leads to static model of Ref.~\onlinecite{SauDasSarma2010,*Alicea2010prb,*OregRefaelvonOppen2010} except for a small SOC component along the magnetic field axis. On the other hand, for static SOC parallel to the IR or the R drivings, the Floquet quasienergies present strong avoided crossings: As it is shown in Sec.\ref{SC:finite} the presence of Majorana states in the corresponding finite systems is discouraged.

\subsection{Finite systems: Floquet Majorana Fermions}
\label{SC:finite}

As discussed in Sec.\ref{SC:infinite} the type of Floquet Majorana fermions expected here, because of the particle-hole symmetry and the absence of $\ve\smeq 0$ localized solutions, have the form given in Eq.~\eqref{EQ:FMajoranaF} with quasienergy $\ve_{\rm FMF} \smeq \ve_{0,n}(\Omega)\smeq(n + 1/2)\hbar\Omega$. Its linear dependence on the driving frequency determines that the mean energy of FMFs is zero, see Eq.~\eqref{EQ:meanEnergy}.
In the following, we apply a numerical method looking for solutions satisfying $\ve\smap \ve_{0,n}(\Omega)$ and $\overline{E}\smap 0$ in finite systems subject to a variety of drivings. Our numerical approach solves the eigenvalue problem for the evolution operator over one driving period, obtaining the quasienergies based on Eq.\eqref{EQ:evolOp}. Spatially inhomogeneous interactions and drivings are easily included by using a tight-binding lattice with lattice spacing $a_0$ and $N$ sites. This is briefly described in the Appendix \ref{AP:A}. As the evolution operator is evaluated from $t\smeq 0$ to $t\smeq T$, its eigenvectors are the instantaneous form of the Floquet states at $t\smeq0$.

\begin{figure}[!b]\begin{center}
\includegraphics[width=0.48\textwidth]{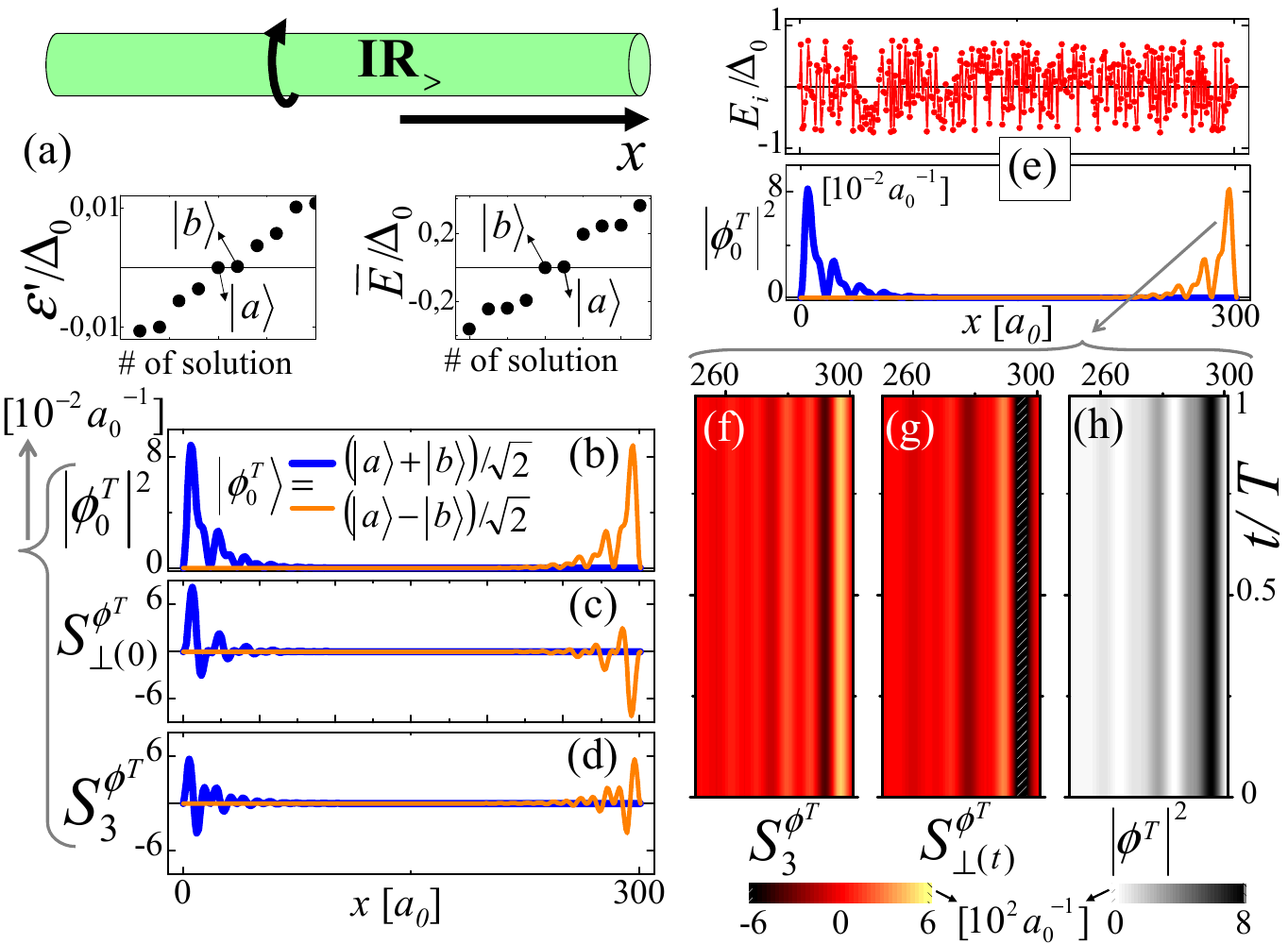}
\vspace{-0.5cm}
\caption{\label{FG:fig5} Floquet Majorana states in a finite sample subject to the IR excitation in the $\Omega>\Omega_{\rm c}$ phase with $\tilde{E}_\mathrm{so}\smeq\alphatilde^2 m^*/(2 \hbar^2)\smeq 0.5\Delta_0$. Reference quantities are $\Delta_0$ and the wavelength $\lambda_\Delta\smeq 2\pi/k_\Delta$, with $k_\Delta\smeq \sqrt{2 m^* \Delta_0} /\hbar$. The lattice spacing is $a_0\smeq 0.0225\lambda_\Delta$ (see text) with sample length  $300 a_0 \smeq 6.75\lambda_\Delta$. We choose $\mu\smeq 2\Delta_0$ such that $\hbar\Omega_{\rm c}\smeq 2\sqrt{5}\Delta_0$, with driving frequency $\Omega\smeq 1.15\Omega_{\rm c}$. (a) Floquet solutions ordered by growing quasienergy (left) and mean energy (right). We plot $\ve'\smequiv \ve\smmi\hbar \Omega/2$ for $0 \le \ve \le\hbar\Omega$. Two edge states, $\ketLR{a}$ and $\ketLR{b}$, are found with $\ve \smap \hbar\Omega/2$ and mean energy $\overline{E} \smap 0$. Isolated Floquet Majorana fermions at each edge, $\ketLR{\phi^T}$, are extracted from the bonding and antibonding combinations of $\ketLR{a}$ and $\ketLR{b}$ (see text). For the two FMF states at $t\smeq 0$ we plot, as a function of the position $x$: (b) The probability density, $\abssqr{\phi_0^T}\smequiv \abssqr{\phi^T (x,t\smeq 0)}$; (c) the electron spin density (ESD) along $\sigma_{\perp(0)}\smeq\sigma_2$, $S_{\perp(0)}^{\phi_0^T}\smequiv S_{\perp(t)}^{\phi^T}(x,t\smeq0)$; and (d) the ESD along $\sigma_3$, $S_{3}^{\phi_0^T} \smequiv S_{3}^{\phi^T}(x,t\smeq0)$. The ESD along $\sigma_{\parallel(0)}\smeq\sigma_1$ (not shown) is zero. (e) We include uncorrelated disorder $E_i\in[-0.75\Delta_0,0.75\Delta_0]$. FMFs are found with probability density similar to that of clean systems shown in (b). FMFs' densities are plotted as a function of time along one driving period in the presence of disorder: (f) $\abssqr{\phi^T(x,t)}$, (g)  $S_{\perp(t)}^{\phi^T}(x,t)$, and (h) $S_{3}^{\phi^T}(x,t)$. The in-plane spin component rotates while the FMF state conserves its location (see text).}
\end{center}
\vspace{-0.6cm}
\end{figure}

We start by studying a clean, finite sample subject to IR driving in the supercritical regime $\Omega>\Omega_{\rm c}$. Bound states at the edges are expected as a result of change in topology since the vacuum is in the topologically trivial phase (see Eq.\eqref{EQ:condition} and the related discussion in Sec.\ref{SC:deriving}). In Fig.~\ref{FG:fig5}(a) we show the resulting quasienergies and mean energies. We sort the solutions according to growing quasienergies (left panel) or to growing mean energies (right panel), finding two solutions which are close to the Majorana conditions
\be
(\ve-\hbar\Omega/2)\!\!\! \mod  \hbar\Omega = 0~,~~~ \overline{E}\smeq 0.
\label{EQ:MajoranaCondition}
\ee
We call these solutions $\ketLR{a}$ and $\ketLR{b}$. Both states have significant probability weight at the two edges. [Similar eigenstates are found with energies $\pm \delta E$ (with $\delta E$ small but different from zero) in a topological superconducting finite sample described by the static model of Ref.~\onlinecite{SauDasSarma2010,*Alicea2010prb,*OregRefaelvonOppen2010}.] These eigenstates are bonding and antibonding combinations of Majorana fermions at the edges produced by finite-size effects. To get rid of the mixing we undo the (anti)bonding as\footnote{An extra phase factor might be needed to decouple the two solutions, this depends on the direction of the SOC at $t\smeq 0$ and/or on the arbitrary phases accompanying each of the computed eigenvectors.}
\be
\ketLR{\phi_\pm^T(t\smeq 0)}=\frac{1}{\sqrt{2}}\left(\ketLR{a}\pm\ketLR{b}\right).
\label{EQ:FMFfinite}
\ee
In this way (see Fig.\ref{FG:fig5}(b)) we obtain two independent Floquet Majorana solutions, $\ketLR{\phi_+^T(t\smeq 0)}$ and $\ketLR{\phi_-^T(t\smeq 0)}$, one on each sample's edge. These are the solutions actually expected on long samples where the overlap between localized states is exponentially small.

For studying the structure of the these solutions we first introduce the electron-hole identity and Pauli matrices $\tau_j$ with $j\smeq 0,1,2,3$. In this way, the operators in the Nambu space can be written as a combination of the operators
\be
\tau_j\sigma_i\equiv \left(\begin{array}{cc}
\tau_{j,11}\sigma_i & \tau_{j,12}\sigma_i \\
\tau_{j,21}\sigma_i & \tau_{j,22}\sigma_i
\end{array}\right),
\ee
where $\tau_{j,\gamma \delta}$ ($\gamma, \delta\smeq1,2$) are the components of the $\tau_j$ matrix such that $\tau_{j,\gamma \delta}\sigma_i$ are $2\times 2$ blocks operating on spin space. The components $\{1,2,3,4\}$ in the Nambu space are associated with $\tau\sigma\smeq\{++,+-,-+,--\}$, respectively, where $\tau$ and $\sigma$ refer to the $\pm 1$ eigenvalues of $\tau_3$ and $\sigma_3$, respectively. Notice that the combination $\tau_e\smequiv(\tau_0+\tau_3)/2$ [$\tau_h\smequiv(\tau_0-\tau_3)/2$] has one single nonzero element, $\tau_{e,11}\smeq 1$ [$\tau_{h,22}\smeq 1$], and it is useful for writing operators associated with the electron [hole] sector only.

For the time-dependent solution $\ketLR{\phi\vphantom{\phi^T}(t)}\smeq {\rm e}^{-\frac{\ci}{\hbar}\ve t}\ketLR{\phi^T(t)}$ we define the instantaneous probability density $|\phi^T(x,t)|^2$ as the sum of the electron and hole probability densities of the associated Floquet state, $\rho^{\phi^T}_{e/h}(x,t)$. These are obtained when computing the expectation value of the identity operator $\hat{\eins}\smequiv \int dx \ketLR{x}\braLR{x}(\tau_e \smpl \tau_h)\sigma_0$:
\bea
\braLR{\phi(t)} \hat{\eins} \ketLR{\phi(t)}=\int_{-\infty}^{\infty} dx \left( \rho^{\phi^T}_{e}(x,t)+ \rho^{\phi^T}_{h}(x,t)\right)  ,~\nonumber~~~~~~~~~~\\
\rho^{\phi^T}_{e/h}(x,t)\equiv \sum_{\sigma,\tau,\tau'}\braket{\left.\phi^T(t)\right|x\tau\sigma} \braket{\tau|\tau_{e/h}|\tau'} \braket{x\tau'\sigma\left|\phi^T(t)\right.}.~~~~~~~~~~\label{EQ:DensP}
\eea
Similarly, related spin densities also evolve in time. The total spin density, $S^{\phi^T}_{i,{\rm tot}}(x,t)$, is the sum of the electron and hole probability densities in the associated Floquet state: $S^{\phi^T}_{i,e/h}(x,t)$. These are obtained by computing the expectation value of  $\hat{S_i}\smequiv \int dx \ketLR{x}\braLR{x}(\tau_e \smpl \tau_h)\sigma_i$, with $i\smeq 1,2,3$ ($\hbar/2$ factors skipped for simplicity):
\bea
\braLR{\phi(t)} \hat{S_i} \ketLR{\phi(t)}=\int_{-\infty}^{\infty} dx \left( S^{\phi^T}_{i,e}(x,t)+ S^{\phi^T}_{i,h}(x,t)\right) ,~~~~~~~~~~~~~\label{EQ:DensS}\\
S^{\phi^T}_{i,e/h}(x,t)\equiv\!\!\!\!\! \sum_{\sigma,\sigma',\tau,\tau'} \!\!\!\!\braket{\left.\phi^T(t)\right|x\tau\sigma} \braket{\sigma,\tau|\tau_{e/h}\sigma_i|\sigma',\tau'} \braket{x\tau'\sigma\left|\phi^T(t)\right.}. \nonumber
\eea
Notice that the densities defined in Eq.\eqref{EQ:DensP} and Eq.\eqref{EQ:DensS} are properties of the physical state since the quasienergy phase factor cancels out. Such cancelation guarantees that the resulting densities are the same irrespective of which shifted version of the Floquet state (see Eq.\eqref{EQ:shifted}) is taken.
\begin{figure*}[!t]\begin{center}
\includegraphics[clip,width=0.85\textwidth]{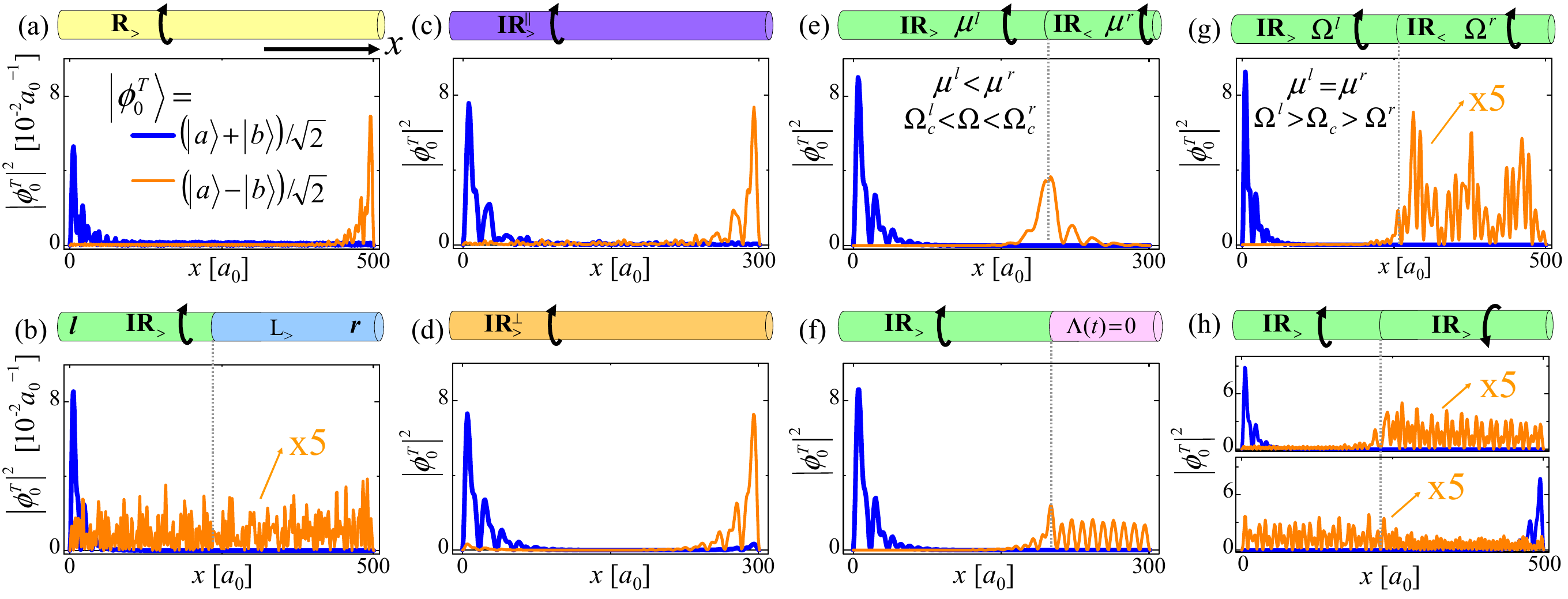}
\vspace{-0.4cm}
\caption{\label{FG:fig6} Floquet Majorana states in finite samples subject to different types of SOC drivings and/or static SOC. All samples include disorder as the one given in Fig.~\ref{FG:fig5}(e). Unless otherwise stated the parameters $a_0$, $\mu$, $\Omega_{\rm c}$, $\Omega\smeq 1.15\Omega_{\rm c}$ and $\tilde{E}_\mathrm{so}$ are those given in Fig.~\ref{FG:fig5}. As in Fig.\ref{FG:fig5}, decoupled solutions are obtained by combining Floquet states $\ketLR{a}$ and $\ketLR{b}$ with $\ve\smap\hbar\Omega/2$ and $\overline{E}\smap 0$. In all cases we plot the probability density $\abssqr{\phi_0^T}\smequiv \abssqr{\phi^T (x,t\smeq 0)}$ of the two (or more) FMF-candidate states. We rescale some curves ($\times5$) to improve visibility. (a) Full sample with R driving. (b) Left ($l$) section of the sample is subject to IR driving whereas the right section ($r$) is L-driven. (c)-(d) IR-driven sample in the presence of a static SOC with $\alpha_0\smeq0.2\alphatilde$ either parallel (c) or perpendicular (d) to the plane of the IR driving. (e) IR-driven sample with inhomogeneous chemical potential on the left and right sections, $\mu^l\smeq 2\Delta_0\smeq \mu^r/2$. The driving frequency is $1.15\Omega^{l}_{\rm c}$ but smaller than the $\Omega^{r}_{\rm c}$. (f) IR-driven sample on the left section in contact with a zero-SOC section on the right. (g) IR-driven sample subject to different driving frequencies on left and right sections, $\Omega_l\smeq 2\Omega_r\smeq 1.15\Omega_{\rm c}$: The left (right) section is in the supercritical (subcritical) regime. (h) Sample subject to counterrotating IR drivings with $\Omega\smeq1.15\Omega_{\rm c}$, clockwise (counterclokwise) on the left (right) section of the sample. Four Floquet states arise in this case.}
\end{center}
\vspace{-0.6cm}
\end{figure*}

As mentioned above, the states $\ketLR{\phi_\pm^T(t)}$ studied in this section [based on Eq.\eqref{EQ:FMFfinite}] satisfy (numerically) the Majorana condition of Eq.\eqref{EQ:MajoranaCondition}. From hereon, we use $\ketLR{\phi^T(t)}$ to refer to any of these two states. In all studied situations (Figs.~\ref{FG:fig5} and \ref{FG:fig6}) we find $|\phi^T(x,t)|^2/2\smeq\rho^{\phi^T}_{e}(x,0)\smeq\rho^{\phi^T}_{h}(x,0)$ and $S^{\phi^T}_{i,tot}(x,0)\smeq 0$ for all $i\smeq1$, $2$ and $3$, namely, $S^{\phi^T}_{i,e}(x,0)\smeq -S^{\phi^T}_{i,h}(x,0)$. Notice this differs from the regular type of quasiparticle states arising in an \emph{s}-wave spin-degenerated BdG Hamiltonian: The latter can not have vanishing $S^{\phi}_{i,tot}(x,0)$ for all three spin directions simultaneously. However, the result coincides with what is expected for a Majorana fermion as its own antiparticle (see Nambu-space representation of Eq.\eqref{EQ:FMajoranaF}). In what follows, we focus on the spin densities for electrons $S^{\phi^T}_{i}(x,t)\smequiv S^{\phi^T}_{i,e}(x,t)$ as it is meaningless to discuss the total spin of a Majorana solution.

In Fig.\ref{FG:fig5}, for both clean and disordered systems subject to IR driving in the $\Omega > \Omega_{\rm c}$ phase, we present the total probability density $|\phi^T(x,t)|^2$ and the electron spin densities (ESDs). In Fig.\ref{FG:fig5}(d) we see that the states at different edges have the same $S^{\phi^T}_{3}(x,0)$ profile. Moreover, these profiles do not change with time, as shown in Fig.\ref{FG:fig5}(f). This is equivalent to that found for the ESD of MFs along the magnetic-field axis in the static system of Ref.~\onlinecite{SauDasSarma2010,*Alicea2010prb,*OregRefaelvonOppen2010} (where the larger the SOC in relation to the Zeeman energy, the smaller such a component\cite{Simon2012}). In our non-magnetic system, the associated direction is normal to the plane defined by the IR driving--- see Eq.\eqref{EQ:ideal}.

Additionally, in Fig.\ref{FG:fig5}(c) we see that the FMF solutions have opposite electron spin in-plane components which change as a function of time, as shown in Fig.~\ref{FG:fig5}(g). We have defined the parallel, $\sigma_{\parallel(t)}$, and the perpendicular, $\sigma_{\perp(t)}$, spin directions with respect to the \emph{instantaneous} SOC axis $\mathbf{\Lambda}_\mathrm{IR}(t)$:
\bese
\bea
\sigma_{\parallel(t)}\equiv \sigma_1 \cos(\Omega t)-\sigma_2 \sin(\Omega t), \\
\sigma_{\perp(t)}\equiv \sigma_2 \cos(\Omega t)+\sigma_1 \sin(\Omega t).
\eea
\eese
We find that the spin density $S^{\phi^T}_{\parallel(t)}(x,t)$ vanishes (not shown) while $S^{\phi^T}_{\perp(t)}(x,t)$ in Fig.\ref{FG:fig5}(d) remains constant.
This indicates that the in-plane electron spin components rotate as a function of time. This rotation is already  expected from Eq.~\eqref{EQ:FMajoranaF} as a mapping of the spin properties of MFs present in the static system of Ref.~\onlinecite{SauDasSarma2010,*Alicea2010prb,*OregRefaelvonOppen2010}, which show a finite ESD \emph{normal} to both the SOC axis and the magnetic-field axis. Such spin component assumes opposite signs for MFs located at opposite edges of the sample,\cite{Simon2012} this is in agreement with our results entirely obtained using Floquet tools for the driven non-magnetic system.

Figure \ref{FG:fig5}(e) presents one realization of uncorrelated disorder incorporated in all remaining results of this section, including time-evolved properties of Floquet states depicted in Figs.~\ref{FG:fig5}(f), \ref{FG:fig5}(g) and \ref{FG:fig5}(h). The disorder is implemented as tight-binding onsite energy fluctuations $E_i\in[-0.75\Delta_0,0.75\Delta_0]$ chosen from an uniform distribution. Its characteristic energy is comparable to $\Delta_0$ and larger than the SOC-driving energy scale, $\tilde{E}_\mathrm{so}$. This site by site uncorrelated disorder avoids the presence of long sections of homogeneous large disturbances driving parts of the system outside of the topological phase being characterized. The robustness of the results under this strong disorder is a signature of their topological origin.

Figure \ref{FG:fig6} shows several finite samples involving different type of drivings or local parameters: All situations include disorder as the above described. There we plot the probability densities for the states $\ketLR{\phi^T(t)}$ at $t\smeq0$ at a supercritical driving frequency $\Omega\smeq 1.15 \Omega_{\rm c}$. Figure \ref{FG:fig6}(a) presents results for R driving. Localized states at the edges have different profiles due to the presence of higher harmonics, evolving as a function of time in contrast to IR driving (not shown). Despite the fact that the non ideal rotation seems to introduce noise in the unpaired FMFs, the very presence of those states given the strong local disorder is remarkable.

On the other hand, systems subject to L  driving (not shown) do not present Floquet states satisfying the Majorana conditions of Eq.~\eqref{EQ:MajoranaCondition}. This is due to the double degeneracy of Floquet states, inconsistent with topologically protected unpaired edge states. Similarly, the presence of partial L driving tends to delocalize Majorana states. In Fig.\ref{FG:fig6}(b) we show a system subject to hybrid IR/L drivings on the left and right of the sample, respectively. The presence of IR driving introduces Majorana-like solutions but only one of them is localized (at isolated edge of the IR section), the partner is delocalized all over the sample due to the L driving.

Figures \ref{FG:fig6}(c) and \ref{FG:fig6}(d) show results for IR driving in the presence of an small static SOC component, either parallel or normal to the plane defined by the driving, respectively. In either case, localized Majorana solutions are found at the sample edges. The normal configuration is less affected by the static SOC, presenting a highly symmetric state distribution in contrast to the parallel configuration. This is consistent with the behavior of the mean energies (for cases $\mathrm{IR}^\parallel$ and $\mathrm{IR}^\perp$ in Fig.\ref{FG:fig4}) for extended solutions with small $k$.

In Fig.\ref{FG:fig6}(e) we present the results for IR driving with inhomogeneous chemical potential leading to position-dependent critical frequencies. The chosen driving frequency is supercritical on the left and subcritical on the right. Majorana states are found at the edges of the left section. This is in full agreement with what is expected by virtue of the mapping to the static topological system of Ref.~\onlinecite{SauDasSarma2010,*Alicea2010prb,*OregRefaelvonOppen2010}. In Fig.\ref{FG:fig6}(f) we show the Floquet Majorana states for the case in which the left section is subject to IR driving ($\Omega>\Omega_{\rm c}$) while the right part is free of any SOC. We find two Majorana states, the one at the interface is delocalized as it penetrates into the unexcited section. This is expected, indeed, since the section with vanishing SOC maps (by applying the global time-dependent spin rotation) into a static superconductor in which the Zeeman energy is \emph{larger} than $\Delta_0$, leading to $E\smeq0$ solutions that are not gapped because the SOC is zero.

In Fig.~\ref{FG:fig6}(g) we have investigated a situation in which the sample is subject to inhomogeneous IR-driving frequencies: $\Omega^{\rm r}$ on the right section and $\Omega^{\rm l}=2\Omega^{\rm r}$ on the left one. In this way Floquet theorem can be applied as we avoid dealing with incommensurable frequencies. We choose a supercritical $\Omega^{\rm l}= 1.15 \Omega_{\rm c}$ such that $\Omega^{\rm r}$ is subcritical. The results are similar to the case of Fig.~\ref{FG:fig6}(f), with a localized Majorana solution on the left and a delocalized state penetrating into the right section.

Finally, we obtained the Floquet solutions for counterrotating IR drivings: Clockwise rotation on the left section and counterclockwise on the right one, see Fig.~\ref{FG:fig6}(h). The situation results interesting since both sides satisfy the supercritical condition $\Omega>\Omega_{\rm c}$. As in all other cases, we have chosen the position of the left/right interface away from the geometrical center of the sample to avoid solutions influenced by symmetry. We find two pairs of states satisfying the Majorana condition. The finite-size offsets to Eq.\eqref{EQ:MajoranaCondition}, $\pm\delta\ve$ and $\pm\delta \overline{E}$, are different for the two pairs. For each pair we find a localized FMF at one edge and a delocalized solution over the opposite side of the sample.

The results discussed along this section, obtained in the presence of disorder, confirms the existence of bound Floquet states at the edges systems subject to IR driving. They are Floquet Majorana fermions arising in absence of magnetic fields when the driving frequency larger than $\Omega_{\rm c}$. When the samples are enlarged, the local probability densities of the delocalized FMF solutions decrease [see Figs.~\ref{FG:fig6}(b), \ref{FG:fig6}(f), \ref{FG:fig6}(g), and \ref{FG:fig6}(h)] whereas the localized solutions remain unaffected.

\section{Conclusions}
\label{SC:conclusions}

Starting from a well-known quantum system hosting MFs for sufficiently large magnetic fields,\cite{SauDasSarma2010,*Alicea2010prb,*OregRefaelvonOppen2010} we derived a non-magnetic scheme sharing the same topological properties thanks to a periodic driving. Our proposal disregards Zeeman coupling: Its role is played by the driving frequency of a rotating SOC axis. An effective breaking of TR symmetry is induced by the definite rotation sense (either clockwise or counterclockwise).
Otherwise, double degeneracy (due to a TR symmetry) would discourage the appearance of a single Majorana solution at the sample's edges.
The experimental realization of this theoretical model would require the ability to control the SOC axis as a function of time. This can be envisioned in quantum wires\footnote{Not only physical \emph{quantum wires} but also quasi 1D-electron gases realized in gate-tunable quantum-well heterostructures.} subject to orthogonal, lateral gates modulating the symmetry of the potential confining a electron system. Alternatively, in cold-atom quantum wires the Rashba coupling could be engineered to change its axis with time.\cite{Cirac2011prl,Zhang2010SOCinColdAtoms} The proposal opens the possibility of non-magnetic platforms for MFs in the field of Floquet topological matter.\cite{Kitagawa2010prb,FloquetTopoInsNatPhys2011,FoaCalvoPastawski2011,FertigPRL2011,KitagawaWhiteQW2011,*Kitagawa2010pra,Cirac2011prl,Liu2012superfluid}

We show that in the case of ideal rotating (IR) driving, the \emph{effective Hamiltonian} over one period of evolution, $\hat{H}_{\rm eff}$, is---apart from a global quasienergy shift---the Hamiltonian of the static system of Ref.~\onlinecite{SauDasSarma2010,*Alicea2010prb,*OregRefaelvonOppen2010} which has Zeeman interaction, an \emph{s}-wave superconducting pairing and SOC. For supercritical driving frequency $\Omega > \Omega_{\rm c}$, (i.e., within the Floquet TSP), unpaired Floquet Majorana fermions are found at the edges of finite samples. While in static systems MFs appear at zero energy, the FMFs in this platform have quasienergy $\hbar\Omega\left(n+1/2\right)$; the existence of FMFs with this quasienergy value was reported in Ref.~\onlinecite{Cirac2011prl} (in a magnetic Floquet system). We notice that these FMFs solutions have \emph{zero mean energy} as a consequence of the linear dependence of FMF-quasienergies on $\Omega$. This fact can be important in the light of Ref.~\onlinecite{Arimondo2012}, where Arimondo \emph{et al.} suggest (with experimental and theoretical support) that the mean energy might determine the occupancy of Floquet states once the memory on the initial condition is lost.\cite{Breuer2000pre} As in the BdG equation the chemical potential lies at zero energy, the reported FMFs could be the zero temperature \emph{highest mean-energy} occupied Floquet states.

We studied a variety of possible drivings based on time-dependent SOC. Rotating SOC drivings are needed for TR-symmetry breaking. For non-ideal SOC rotations, our numerical results show that higher harmonics can degrade the formation of FMFs. Similar degradation is found in the case of an IR driving coexisting with a static SOC component within the driving plane. We demonstrate the great deal of possibilities for FMFs at edges and interfaces by exploring only a handful of examples.

For finite samples, our numerical simulations show that the FMFs appear even in the presence of disorder. This is remarkable, considering that quasienergy bands do not appear to be protected by a quasienergy gap: At larger values of $k$ the gap closes. However, the underlying energy ``gap" protecting the Floquet Majorana solutions can be extracted from $\hat{H}_{\rm eff}$ by studying the topology of its gapped phases.\cite{Kitagawa2010prb} As mentioned above, in the case of IR driving this can be done analytically: One finds that the gap protecting the FMFs is the same as the one for the quantum wire platforms of Ref.~\onlinecite{SauDasSarma2010,*Alicea2010prb,*OregRefaelvonOppen2010} after replacing the Zeeman energy in those models by $\hbar\Omega/2$.\footnote{For example, when $\tilde{E}_\mathrm{so}\ll \hbar\Omega/2$ then the effective \emph{p}-wave protecting gap would scale as $\Delta_0 \tilde{E}_\mathrm{so}/\hbar\Omega$.}

Finally, we expect that tunneling probing (with a normal lead) of one edge of a long sample in the Floquet TSP shall lead to peaks in the differential conductance $dI/dV$ at bias voltages $V_\pm\smeq\pm\hbar\Omega/2$. The peaks heights are determined by the ratio between $\hbar\Omega/2$ (i.e., the effective Zeeman energy) and the spin-orbit coupling strength. The peaks are not present in the topologically trivial phase (subcritical $\Omega<\Omega_\mathrm{c}$), though a background with some structure is visible in both phases (including pumping effects,\cite{MoskaletsButtiker2002pumping,Braatas2004prl,Arrachea2005} as a finite current for zero bias due to the rotating-SOC). These qualitative conclusions are justified within a picture that makes use of the rotating frame transformation (where the driven part of the sample is mapped to the static topological phase, having an unpaired MF at its edge at $E\smeq 0$), introducing a spin-dependent shift $\pm \hbar\Omega/2$ in the Fermi energy of normal probes.\cite{ChenNikolic2009}
These shifts are introduced to conserve the number of electrons in the static lead before and after the transformation (notice that the effective magnetic field splits the $up$ and $down$ bands) as the ultimate goal is to study the transport from the normal lead to the driven system.\footnote{Notice that in this work we have applied the time-dependent unitary transformation of Eq.\eqref{EQ:UR} to the states as a mathematical tool to obtain the solutions of the time-dependent BdG-Schroedinger equation dictated by $\hat{H}(t)$. In such mathematical framework the chemical potential $\mu$ appearing in $\hat{H}(t)$ is not modified at all when applying the transformation to the states.}

As a pending issue, it stands out an extensive study on the appropriate observables for detecting the Floquet TSP. Results beyond the IR case would require a Floquet-Keldysh approach.\cite{KohlerReview2005,FoaCalvoPastawski2011,Mahfouzi2012}  Another line includes developing schemes to manipulate localized FMFs (the ones reported here or those reported in Ref.\onlinecite{Cirac2011prl}) for quantum information purposes: Quantum memory, braiding, etc. In all cases, an understanding of the Floquet-state occupancy, a problem that falls within the complex subject of statistical mechanics for driven systems,\cite{Hone2009,Ketzmerick2010} appears to be crucial, as does studying the role of time-dependent noise and noise mitigation schemes. Here, we have pointed out that interesting physical properties can be expected by developing the ability to manipulate the SOC axis as a function of time. This motivates the search for alternative platforms achieving this ability (e.g., new sample designs using known materials), including situations where superconducting pairing is absent.

\acknowledgments

We acknowledge useful discussions with A. Doherty, K. Flensberg, A. Lobos and J. M. Taylor. AAR acknowledges support from the Australian Research Council Centre of Excellence scheme CE110001013, from ARO/IARPA project W911NF-10-1-0330, and the hospitality of University of Seville. DF acknowledges support from the Ram\'on y Cajal program, from the Spanish Ministry of Science and Innovation's projects No. FIS2008-05596 and FIS2011-29400, and from the Junta de Andaluc\'ia's Excellence Project No. P07-FQM-3037.

\appendix
\section{Solving the Floquet problem}
\label{AP:A}
Floquet theorem is the equivalent to the Bloch theorem for periodic driving instead of spatial periodic potentials, where the quasienergies of Floquet states play the role of momenta of Bloch states. The equivalence of solutions
belonging to different Brillouin zones in momentum is one of the main signatures of Bloch systems. Similar properties are shared by Floquet systems in the quasienergy axis. This can be seen from Eq.~\ref{EQ:physical}, after noticing that the physical states $\ketLR{\phi_{a}(t)}$ remain unchanged by the substitution
\be
\varepsilon_{a} \rightarrow \varepsilon_{a} + n \hbar \Omega~,~~
\ketLR{\phi^T_{a}(t)} \rightarrow \ket{\phi^{T,n\mathrm{-shift}}_{a}(t)}\equiv{\rm e}^{\ci n \Omega t}  \ketLR{\phi^T_{a}(t)}.
\label{EQ:shifted}
\ee
Namely, the Floquet state $\ket{\phi^{T,n\mathrm{-shift}}_{a}(t)}$ has shifted Fourier components, therefore, for each physical state there are infinite Floquet QESs that are nonorthogonal, with $\braket{\phi^{T}_{a}(t)| \phi^{T,n\mathrm{-shift}}_{a}(t)}= {\rm e}^{\ci n \Omega t}$. However, orthogonality is recovered by defining the inner product of Floquet QESs as the standard inner product averaged over one driving period,
\be
\braket{\braket{\phi^{T}_{a}(t)| \phi^{T}_{b}(t)}}\equiv\frac{1}{T}\int_0^T  \braket{\phi^{T}_{a}(t')| \phi^{T}_{b}(t')} dt'=\delta_{a,b}.
\ee
\begin{widetext}
To retain only one QES for each physical state it is sufficient to choose any range of quasienergies of length $\hbar \Omega$--- such as $[\varepsilon,\varepsilon\smpl \hbar \Omega)$--- and discard all the QESs outside this region.

The time-dependent Eq.~(\ref{EQ:floquetQES}) for the Floquet states still needs to be solved. One possibility is to compute the evolution operator and solve Eq.~(\ref{EQ:evolOp}). We do this for the case of a finite piece of wire.  Alternatively, we can switch to Fourier representation by rewriting the Hamiltonian and the Floquet states as
\be
\hat{H}(t)=\sum_{n=-\infty}^{\infty} {\rm e}^{-\ci \Omega n t} \hat{H}^{(n)}~,~~\ketLR{\phi^T_{a}(t)}=\sum_{n=-\infty}^{\infty} {\rm e}^{-\ci \Omega n t} \ketLR{\phi^{(n)}_{a}}.
\label{EQ:HtFourier}
\ee
By direct substitution in Eq.~(\ref{EQ:floquetQES}) we find
$\sum_{m}\left(\hat{H}^{(m)}- n \hbar \Omega \delta_{0,m}\right)\ketLR{\phi^{(n-m)}_{a}} = \varepsilon_a  \ketLR{\phi^{(n)}_{a}}$, which in matrix representation reads
\begin{equation}
 \left(
\begin{array}{ccccc}  \ddots & \vdots  &  \vdots  & \vdots   &  \Ddots \\ \cdots & \left(\hat{H}^{(0)} + \hbar\Omega\right) & \hat{H}^{(-1)} &\hat{H}^{(-2)}& \cdots
\\ \cdots   &  \hat{H}^{(1)} & \hat{H}^{(0)} & \hat{H}^{(-1)} & \cdots
 \\ \cdots & \hat{H}^{(2)}& \hat{H}_{(1)} & \left(\hat{H}^{(0)} -\hbar \Omega\right)&  \cdots \\ \Ddots & \vdots  &  \vdots  & \vdots   &  \ddots
\end{array} \right) \left(
\begin{array}{c} \vdots  \\  \ket{\phi^{(-1)}_{a}} \\ \ket{\phi^{(0)}_{a}} \\ \ket{\phi^{(1)}_{a}} \\ \vdots
\end{array} \right) = \varepsilon_a \left(
\begin{array}{c} \vdots  \\  \ket{\phi^{(-1)}_{a}} \\ \ket{\phi^{(0)}_{a}} \\ \ket{\phi^{(1)}_{a}} \\ \vdots
\end{array} \right).
\label{EQ:floqMatrix}
\end{equation}
\end{widetext}
This configures a time-independent infinite dimensional eigenvalue problem where $\ve_a$ and $\ketLR{\phi^{(n)}_{a}}$ are unknown. Whenever possible, we solve this analytically provided some approximations are introduced, by comparing $\hbar\Omega$ with other energy scales appearing in the Fourier components $\hat{H}^{(n)}$ of Eq.~(\ref{EQ:floqMatrix}). For dealing with general situations we resort to exact numerical approaches.

For the infinite systems of Sec. \ref{SC:infinite}, we extend the numerical method for the calculation of dispersion relations in spatial lattices given by Ando in Ref.~\onlinecite{Ando89}. This is performed by discretizing time along one driving period. For each temporal ``site" $i$ located at $t_i= i \delta t$ the local Hamiltonian $\hat{H}_k(t_i)$ (see Eq.\eqref{EQ:floquetQESk}) acts on the Nambu spinor $\varphi_{k}(t_i)$. The discretized version of the term $-\ci\hbar \frac{d}{dt}$ in Eq.\eqref{EQ:floquetQESk} provides a hopping term between first-neighbor sites. The discretized $H_{\rm F}^k$ can be thought as an effective spatial (in the ``sites" $t_i$) periodic lattice Hamiltonian in Nambu space. The momentum associated with the motion on the lattice $t_i$ is the Floquet quasienergy, $\varepsilon$. Therefore, the Bloch solutions in this effective lattice are the Floquet ones in the time-dependent system: By imposing Eq.\eqref{EQ:floquetQESk} one obtains $\varepsilon$ as the variable that would be the momentum in a spatially periodic lattice. The technical details of the extended Ando method for Floquet systems will be presented elsewhere. The results are equivalent to those found by computing the evolution operator over one period of the driving frequency and then solving the eigenvalue problem of Eq.\eqref{EQ:evolOp}.

In the case of finite size samples we compute the evolution operator within a tight-binding model. We divide the period $T$ in $120$ time slices during which the driving is assumed to be constant. Due to the finite sample size, the system must be treated in real space: We use the customary finite-differences passage from the continuos to a tight-binding model with $N$ sites.\cite{ReynosoAndreev2012} The wavefunction at each site is represented by a Nambu spinor and the pairing potential enters as an on-site coupling between the electron and hole sectors. The evolution operator becomes a $4N\times4N$ matrix, $\hat{U}_T(t)$. The lattice spacing $a_0$ is chosen sufficiently small to turn the kinetic term [second order spatial derivative, now a hopping $t_{\rm h}\smeq \hbar^2/(2m^*a_0^2)$ between neighbor lattice sites] into the largest energy scale in the system. The SOC [proportional to first-order spatial derivatives] leads to nontrivial spin-hopping terms with amplitudes given by the components of the vector $\left( \mathbf{\Lambda}(t) + \mathbf{\Lambda}^{(0)} \right)/(2 a_0)$. As usual, the tight-binding model has the flexibility to deal with any site-dependent perturbation as local disorder or inhomogeneous driving, both discussed in Figs.~\ref{FG:fig5} and \ref{FG:fig6}.

\end{document}